\documentclass{egpubl}
\usepackage{egsr2024}
\usepackage{subfiles}
\usepackage[T1]{fontenc}
\usepackage{dfadobe}

\biberVersion
\BibtexOrBiblatex
\usepackage[backend=biber,bibstyle=EG,citestyle=alphabetic,backref=true]{biblatex} 
\electronicVersion
\PrintedOrElectronic
\ifpdf \usepackage[pdftex]{graphicx} \pdfcompresslevel=9
\else \usepackage[dvips]{graphicx} \fi

\usepackage{egweblnk} 

\usepackage{caption}
\usepackage{graphicx}
\usepackage{subcaption}
\usepackage{amsmath}
\usepackage{amssymb}
\usepackage{placeins}
\usepackage{xspace}

\newcommand{\sysName}{\textsf{ReflectanceFusion}\xspace}

\title[\sysName: Diffusion-based text to SVBRDF Generation]%
      {\sysName: Diffusion-based text to SVBRDF Generation}

\author[Bowen Xue \& Giuseppe Claudio Guarnera\&Shuang Zhao\&Zahra Montazeri]
{\parbox{\textwidth}{\centering Bowen Xue$^{1}$
        and Giuseppe Claudio Guarnera$^{2}$ and Shuang Zhao$^{3}$and Zahra Montazeri$^{1}$   
        }
        \\
{\parbox{\textwidth}{\centering 
$^1$University of Manchester, UK\\$^2$University of York, UK\\ $^3$University of California, Irvine, USA}
}
}
\begin{document}
\teaser{
 \includegraphics[width=1\linewidth, trim=0 0 0 0, clip]{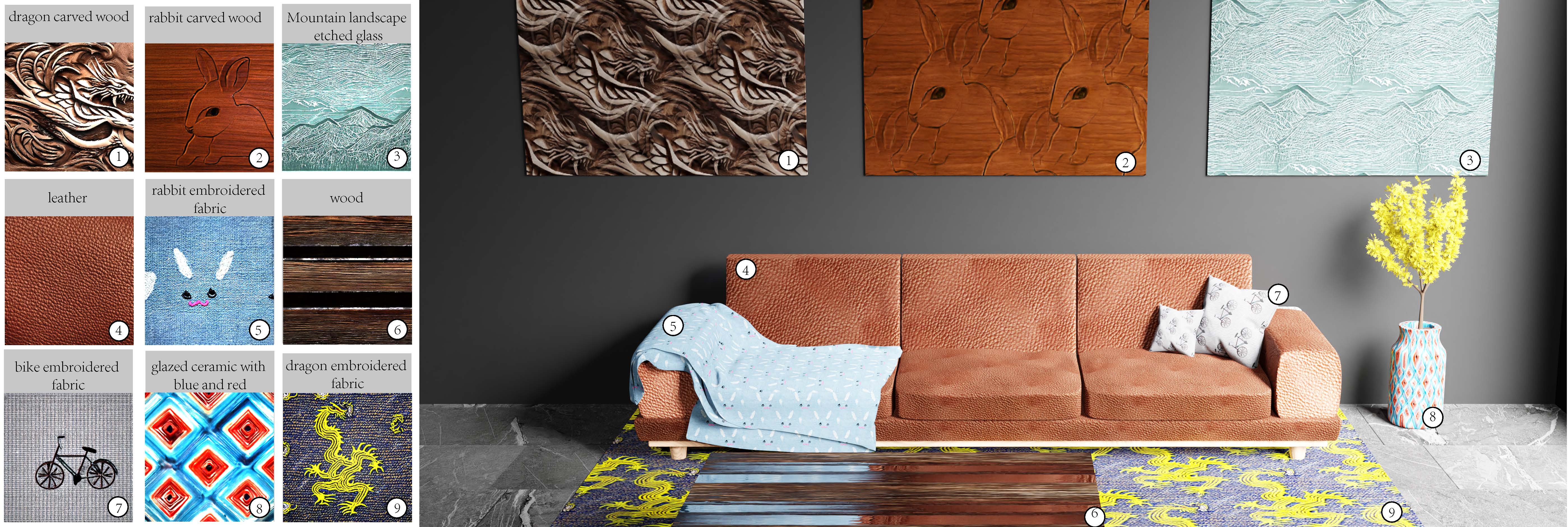}
\caption{Given a textual description of the desired material appearance, our model generates plausible Spatially-Varying Bidirectional Reflectance Distribution Function (SVBRDF) parameter maps. These maps can accurately represent a wide range of materials. Unlike static images, SVBRDFs enable relighting, and editing and can be applied on any object surface. Here we show rendered images under both side and environment lighting to demonstrate the adaptability of our model to diverse lighting configurations.
}
 \centering

\label{fig_teaser}
}

\maketitle
\begin{abstract}
We introduce \sysName (Reflectance Diffusion), a new neural text-to-texture model capable of generating high-fidelity SVBRDF maps from textual descriptions. Our method leverages a tandem neural approach, consisting of two modules, to accurately model the distribution of spatially varying reflectance as described by text prompts. Initially, we employ a pre-trained stable diffusion 2 model to generate a latent representation that informs the overall shape of the material and serves as our backbone model. Then, our ReflectanceUNet enables fine-tuning control over the material's physical appearance and generates SVBRDF maps. ReflectanceUNet module is trained on an extensive dataset comprising approximately 200,000 synthetic spatially varying materials. Our generative SVBRDF diffusion model allows for the synthesis of multiple SVBRDF estimates from a single textual input, offering users the possibility to choose the output that best aligns with their requirements. We illustrate our method's versatility by generating SVBRDF maps from a range of textual descriptions, both specific and broad. Our ReflectanceUNet model can integrate optional physical parameters, such as roughness and specularity, enhancing customization. When the backbone module is fixed, the ReflectanceUNet module refines the material, allowing direct edits to its physical attributes. Comparative evaluations demonstrate that ReflectanceFusion achieves better accuracy than existing text-to-material models, such as Text2Mat, while also providing the benefits of editable and relightable SVBRDF maps.


\end{abstract}  
\section{Introduction}
Reproducing the visual appearance of material reflectance is challenging---especially for materials with detailed spatially varying variations. Conventionally, authoring spatially varying reflectance requires the use of professional software (e.g., Substance) and can be highly time-consuming.
The most promising recent solutions leverage machine learning to generate images corresponding to target text prompts. Existing generative models primarily produce static images under a specific type of lighting, lacking the editability of materials such as SVBRDF maps. Consequently, the rendered images have to be used as-is, with no control over the material.

To tackle this limitation, we formulate SVBRDF estimation as a diffusion task conditioned on an input text description. Our objective is to develop a model that facilitates text-to-SVBRDF conversion, substantially simplifying the complexity of SVBRDF design. Existing diffusion-based models rely on pre-trained, large-scale image diffusion models to sample the distribution of natural images. However, the distribution of SVBRDFs differs significantly from natural images. Therefore, we introduce a novel generative diffusion model tailored for spatially varying materials. We introduce an initial diffusion model to produce an overall appearance and synthesize the input text into a latent space. The output is then passed to a second phase, named ReflectanceUNet, that estimates SVBRDF parameter maps consisting of 10 channels (i.e. normals, diffuse and specular albedo, and specular roughness). Our tandem approach allows us to first generate a rough estimate of the appearance and then refine it in the second stage, governed by physical parameters. 

Furthermore, training diffusion models typically require a significantly larger training set than conventional neural networks. We use the first phase as pre-trained Standard Diffusion 2, serving as a backbone model. To train our second phase, the SVBRDF diffusion model, we supplement the INRIA synthetic SVBRDF dataset \cite{deschaintre2018single} with the UBO dataset \cite{merzbach2020bonn}, amounting to about 200,000 unique training exemplars. We employ Euler as our scheduler and v-prediction for our prediction type, meaning our diffusion model outputs predicted velocity. The loss function used is the root mean square error function. After obtaining the latent representation from the first phase, it goes through VAE and VGG, added together as a connecting module, and fed into the second phase which is a diffusion module, then outputs SVBRDF maps.

Concretely, our contributions include:
\begin{itemize}
    \item Text-to-Texture Pipeline: We have developed a new pipeline that translates textual descriptions into detailed SVBRDF maps, enabling precise and customizable representations of material appearances.
    \item ReflectanceUNet: Our new diffusion-based network generates complex reflectance properties.
    \item Dual-Phase Architectural Strategy: Our method is a two-phase approach, initiating with a backbone network for foundational texture generation, subsequently refined by our ReflectanceUNet for superior detail and accuracy.
\end{itemize}
\section{Related Works}
\subsection{Material Properties Acquisition and Generation}

The SVBRDF \cite{nicodemus1965directional} is a surface reflectance model that characterizes how light reflects off non-homogeneous, opaque surfaces, capturing spatially varying properties such as shininess and texture at different points on the material. Traditionally, acquiring SVBRDF involves documenting the variations in a material's appearance under various lighting and viewing conditions through extensive photographic processes \cite{guarnera2016}. Such image-based approach necessitates high-quality imaging equipment and controlled lighting setups, capturing images from multiple angles and under varied lighting conditions to fully represent the material's reflective behavior \cite{merzbach2020bonn}. In traditional methods, acquiring material properties necessitates expensive and complex professional hardware, such as camera domes and computer-controlled robots, often making it unaffordable for individual artists and forcing them to rely on existing material libraries.  

In the context of material property acquisition, the Bidirectional Texture Function (BTF) \cite{dana1999} merits discussion. Like SVBRDF, BTF captures the texture's appearance under varying lighting and viewing conditions but is tailored for different types of surfaces, as it accounts for effects arising from small-scale geometry, such as inter-reflection and self-shadowing. BTF datasets, such as the UBO dataset \cite{UBO2014}, require specialized equipment and extended collection time, demanding significant storage and retrieval resources. While some studies \cite{xue2023hierarchical} \cite{kuznetsov2021neumip} have successfully compressed BTF using neural networks, achieving good results and speed, they lack generalizability as each material requires training a separate model. Moreover, many methods employ analytical and procedural appearance models \cite{montazeri2020practical, Zhu2023hierarchical}, suitable for complex materials like fabric. Yet, these models are costly and necessitate detailed implementations specific to material types.

 With the advent of neural approaches, predicting SVBRDF from images has been significantly simplified, reducing the number of required images to sometimes just a few, a pair, or even a single one, streamlining material acquisition. Recent works\cite{guo2021highlight,zhou2021adversarial,guo2020materialgan,vecchio2021surfacenet} utilized Generative Adversarial Networks (GANs) for effective image-to-material property transformation. Despite these advancements, accurately replicating materials with complex textures remains a challenge. Sartor et al. \cite{sartor2023matfusion} demonstrate new possibilities by generating SVBRDF from images using diffusion models. Although methods based on procedural material synthesis tools, such as \cite{hu2019novel,shi2020match}, can produce high-quality material maps, they lack scalability. Text2Mat \cite{guo2023text2mat} facilitates the creation of materials from the text but is limited to simpler materials and cannot generate materials with complex meanings.
\begin{figure*}[t]
    \centering
    \includegraphics[width=1\textwidth]{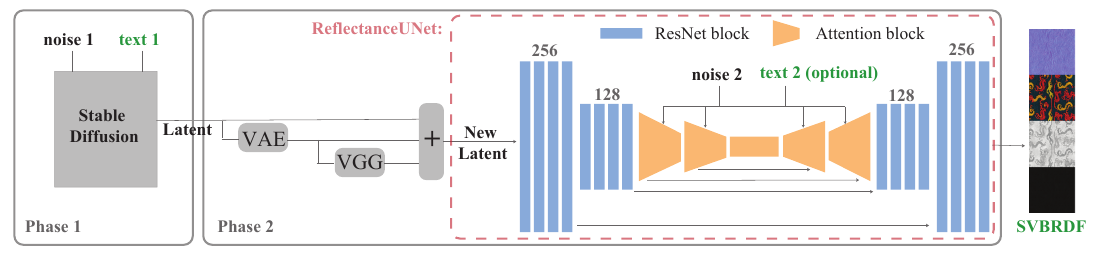}
    \caption{Our dual-phase architecture. The system consists of two parts: Stable Diffusion 2.0 and a custom UNET. The former processes Gaussian noise, text, and time (t), while the latter, a diffusion network, outputs ten-channel SVBRDF. Inputs to the second part include the first's output, VAE and VGG base network contributions, material properties (specularity, roughness), and time step (t).}
    \label{fig_architecture}
\end{figure*}

\subsection{Text to Image Generation}
Variational AutoEncoders (VAE) \cite{kingma2013auto} and GAN \cite{goodfellow2020generative} have been used for image generation tasks, but Diffusion models \cite{dhariwal2021diffusion, ho2020denoising} stand out by producing high-fidelity and diverse images through a multi-step denoising process. In contrast, VAEs and GANs typically generate images in a single step. VAEs often yield blurry images due to overlapping latent code distributions and the averaging effect in pixel reconstruction, while GANs face challenges like mode collapse and difficulties in training convergence.

Diffusion models represent a new generation of generative models that add and then remove noise from images in stages. Unlike traditional networks like VAE and GAN, they operate in lower-dimensional latent spaces to reduce computational load. Latent Diffusion \cite{rombach2022high} employs cross-attention \cite{hao2017end} for multimodal data processing, enhancing model flexibility. Stable Diffusion, built on Latent Diffusion, is optimized for text-to-image tasks and improves generalization through extensive training. It comprises a VAE encoder compression module that encodes images into latent space and a conditional generation module that denoises and diffuses in this space to produce images matching text conditions.

Stable Diffusion XL (SDXL) \cite{podell2023sdxl} is a deep learning model based on the latent diffusion architecture, consisting of two main parts: the Base model and the Refiner model. The Base model handles the overall composition of the image. In contrast, the Refiner model is dedicated to adding finer visual details to further enhance image quality, representing a two-stage diffusion process where each part has a distinct function.

The recent models, Stable Diffusion 3 \cite{esser2024scaling} and Sora (Video generation models as world simulators) \cite{videoworldsimulators2024}, utilize the same underlying architecture, MMDiT (Multimodal Diffusion Transformer), which employs separate sets of weights for image and language representations. The MMDiT architecture uses two independent weight sets for text and image modalities, merging the sequences of both modalities in the attention mechanism. This allows each representation to operate in its own space while considering the other. Stable Diffusion achieved improved results over its predecessors, while Sora demonstrated remarkable text-to-video generation capabilities.
%

\newcommand{\myImageGroup}[2]{
    \begin{minipage}[h]{0.21\textwidth}
        \centering
        \includegraphics[width=\linewidth, trim={2cm 3cm 2cm 1cm},clip]{image3/run_#1/tilepm2noralnew.png}
        \input{image3/run_#1/prompt.txt}
    \end{minipage}
    \begin{minipage}[h]{0.105\textwidth}
        \centering
        \includegraphics[width=\linewidth]{image3/run_#1/00.png}
        \includegraphics[width=\linewidth]{image3/run_#1/08.png}
        \includegraphics[width=\linewidth]{image3/run_#1/renderimage1d.png}
        
        \hfill
    \end{minipage}
}

\newcommand{\myImageGroupb}[2]{
    \begin{minipage}{.33\textwidth}
        \centering
        \includegraphics[width=.325\linewidth]{image3/run_#1/00.png}
        \includegraphics[width=.325\linewidth]{image3/run_#1/04.png}
        \includegraphics[width=.325\linewidth]{image3/run_#1/08.png}
        \includegraphics[width=0.99\linewidth]{image3/run_#1/renderimage1d.png}

        \input{image3/run_#1/prompt.txt}
    \end{minipage}
}
\newcommand{\myImageGroupmetal}[2]{
    \begin{minipage}{.33\textwidth}
        \centering
        \includegraphics[width=.325\linewidth]{image3/run_#1/00.png}
        \includegraphics[width=.325\linewidth]{image3/run_#1/04.png}
        \includegraphics[width=.325\linewidth]{image3/run_#1/08.png}
        
        
        \includegraphics[width=0.99\linewidth]{image3/run_#1/renderimage1d.png}
        
        \input{image3/run_#1/prompt.txt}
    \end{minipage}
}

\section{Our Method}
\subsection{Architectural Design}
Through experimentation, we found that employing a single diffusion model does not simultaneously maintain the original diffusion model's generalization capabilities and meet the high-quality SVBRDF generation requirements. Given the relatively limited SVBRDF dataset (\textasciitilde200,000 samples), to ensure the model's generality, we aimed to maximally preserve the versatility of Stable Diffusion \cite{rombach2022high}. The pipeline design was inspired by Stable Diffusion XL \cite{podell2023sdxl} and other works that use diffusion for generating images and textures. We designed a serial network architecture composed of two diffusion modules, as shown in Fig \ref{fig_architecture}, with the first part being Stable Diffusion 2.0 and the second part our custom-designed U-NET followed by a connecting module. Due to the lack of a sufficiently large dataset to train the SVBRDF VAE structure and the lack of a dataset, we opted for Diffusion instead of latent Diffusion, a network that directly outputs results without an encoder and decoder. This is because latent diffusion operates on denoising in the latent space, requiring a corresponding VAE network to extract the SVBRDF's latent features. However, the limited size of SVBRDF datasets does not provide enough data to effectively train a VAE architecture. In contrast, direct diffusion models denoise in the noise space rather than the latent space, alleviating the issue of having an insufficient amount of data for training. This design choice allows us to maximally retain the original Stable Diffusion's generalization capability. Adopting a serial pipeline with two Diffusion modules instead of one retains the diffusion model's ability to represent appearances. Our dual-phase sequential structure maintains the appearance generation capacity of Stable Diffusion while integrating a second Diffusion module focused on physical parameters, ensuring excellent generalizability and the production of realistic, controllable SVBRDFs.

The first part of the network employs Stable Diffusion 2.0, primarily responsible for denoising the latent appearance aspect. It takes Gaussian noise, textual description, and time step (t) as inputs. The second phase is a diffusion network designed followed by a connecting module to output ten channels, focusing on physical parameters and SVBRF maps. The output of the first diffusion process, augmented with values from the VAE and the VGG base network \cite{simonyan2014very}, along with material properties such as specularity and roughness, and the time step (t), are fed into the second diffusion model. This phase generates a ten-channel output, comprising normals, diffuse albedo, and specular reflection (each with three channels), and one channel for roughness. 

\subsection{Our ReflectanceUNet}

Our network design draws on the structures proposed in previous works \cite{dhariwal2021diffusion,rombach2022high,sartor2023matfusion}, as depicted in Fig \ref{fig_architecture}. Our U-Net follows an encoder-decoder architecture, with both the input and output resolutions set to 256. The output features ten channels, representing normal map, diffuse albedo, roughness, and specular components. To reduce computational demands, the network blocks sized 256 and 128 only contain ResNet blocks; from the 64-sized block onwards, all blocks incorporate attention mechanisms. Each trapezoidal-shaped attention block consists of two ResNet layers and two spatial transformers, ending with a convolutional layer to adjust the output size. The sequence is ResNet layer-spatial transformer-ResNet layer-spatial transformer. Therefore, the input for the leftmost trapezoidal attention block is sized 64, and the output is sized 32, with the trapezoidal shape indicating that the output layer size is halved. The central square-shaped attention layers consist of multiple sequences of ResNet layers and spatial transformers in an alternating pattern. Each attention block has Skip Connections present between the U-Net's encoder and decoder. 

We have opted for Euler as our scheduler and v-prediction \cite{salimans2022progressive} for our prediction type. V-prediction predicts velocity instead of noise. Let \( v \) be the velocity vector, which can be estimated from the model-learned parameters \(x\) and noise \( \epsilon \) to estimate: \( V = \alpha_t \epsilon - \sigma_t x \).
V-prediction helps decrease the variance in generated samples, mitigating issues with limited training data. Additionally, it simplifies the training of diffusion models, resulting in a more stable and interpretable training process. Parameters for roughness and specular are flexible, allowing for inclusion or exclusion as needed. The optimization objective function is defined as follows: 
\begin{equation}
    \mathbb{E}_{t, x \sim p_{\text{data}}(x), \epsilon \sim \mathcal{N}(0, I)} \left\| \tilde{v}_{\theta}(t, \mathbf{z}_t, c) - \mathbf{v} \right\|_2^2
    \label{e_obj}
\end{equation}
where \( \tilde{v}_{\theta} \) is the output of the diffusion model.

\begin{figure}[b]
    \centering
    \begin{minipage}[b]{.02\textwidth}
        \rotatebox{90}{\hspace*{1em} rabbit carved wood \hspace*{5em} wood}
    \end{minipage}%
    \begin{minipage}[t]{.44\textwidth}
        \begin{subfigure}{.32\textwidth}
            \caption*{(a) VAE only}
            \includegraphics[width=\linewidth]{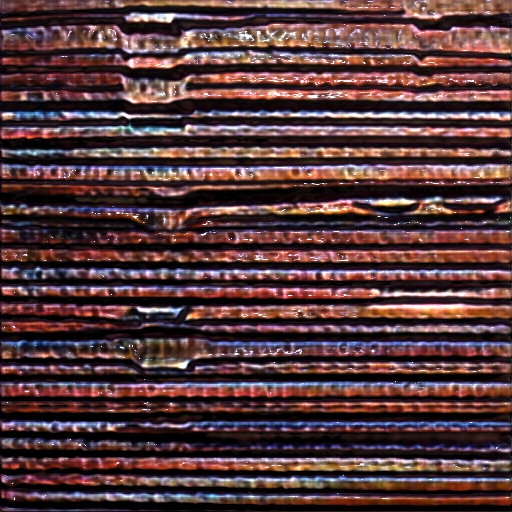}
            \includegraphics[width=\linewidth]{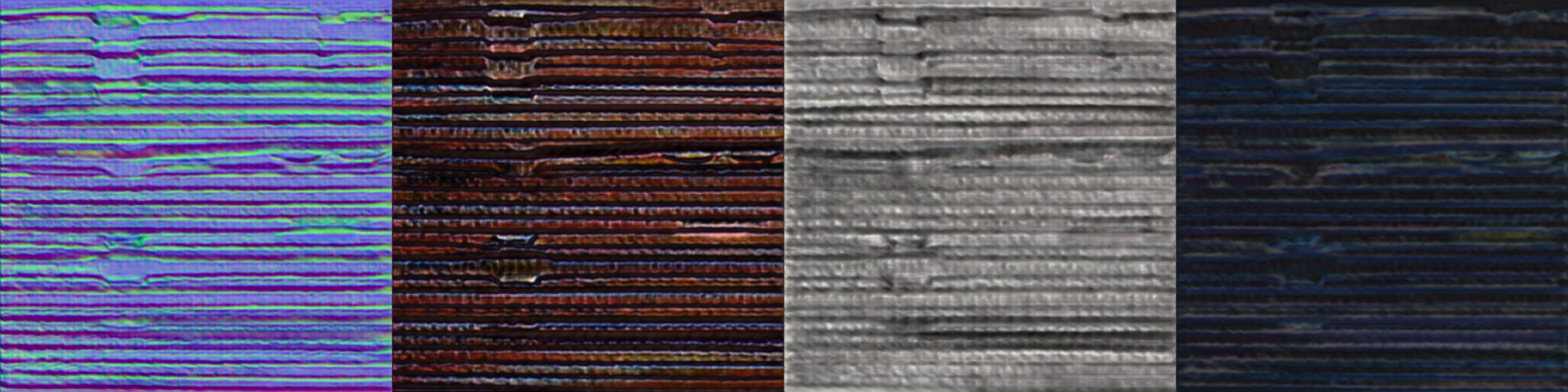}
            \includegraphics[width=\linewidth]{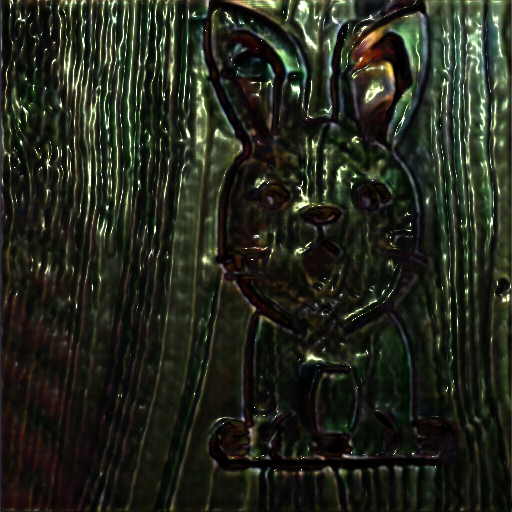} 
            \includegraphics[width=\linewidth]{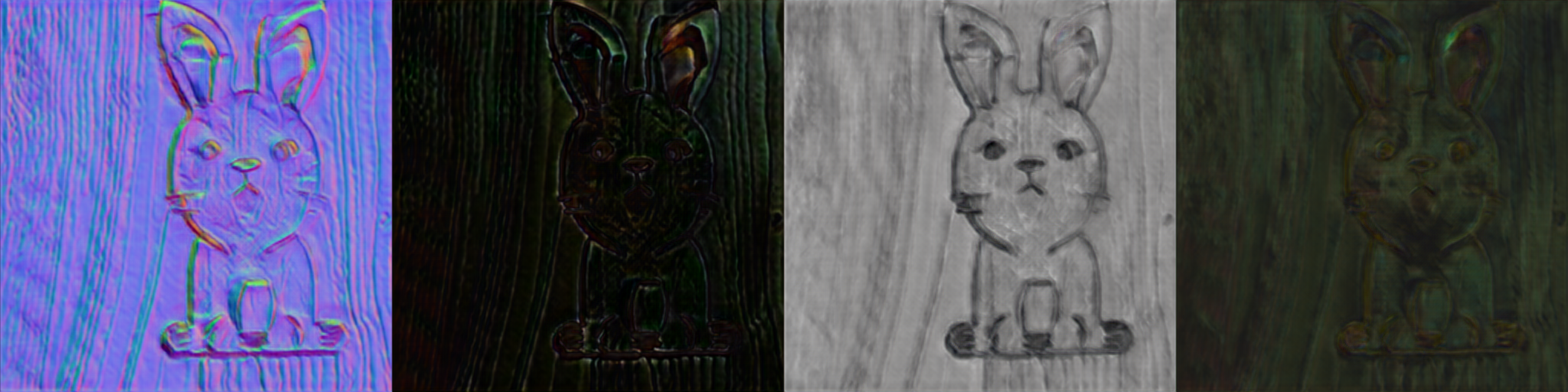}       
        \end{subfigure}
        \begin{subfigure}{.32\textwidth}
            \caption*{(b) Single diffusion}
            \includegraphics[width=\linewidth]{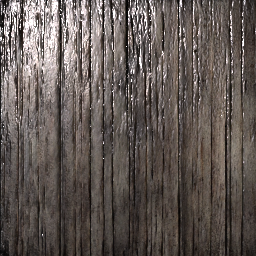}
            \includegraphics[width=\linewidth]{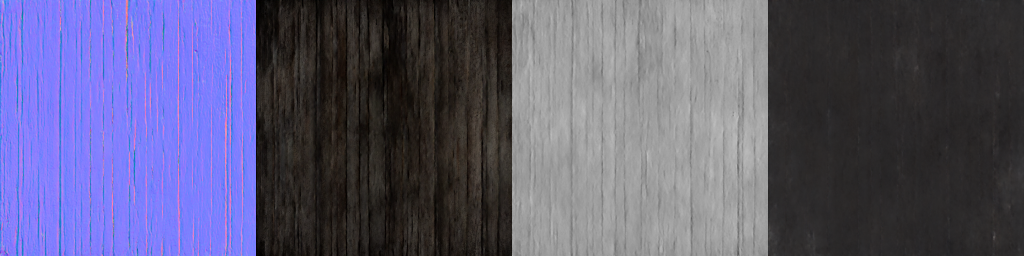}
            \includegraphics[width=\linewidth]{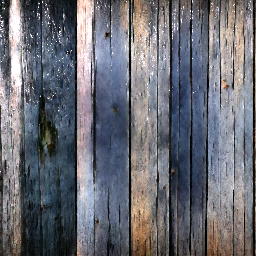} 
            \includegraphics[width=\linewidth]{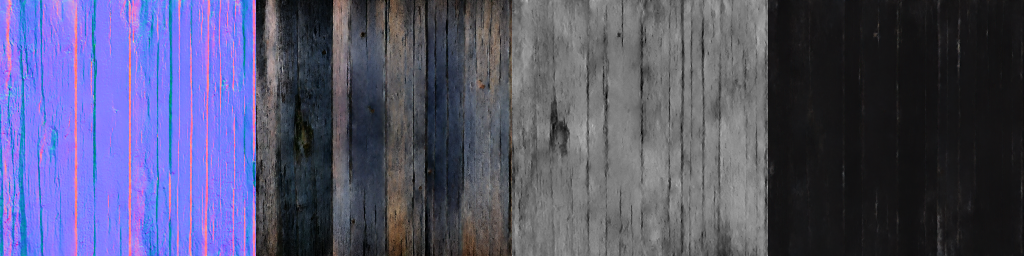} 
        \end{subfigure}
        \begin{subfigure}{.32\textwidth} 
        \caption*{(c) Ours}
            \includegraphics[width=\linewidth]{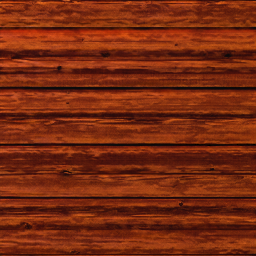}
            \includegraphics[width=\linewidth]{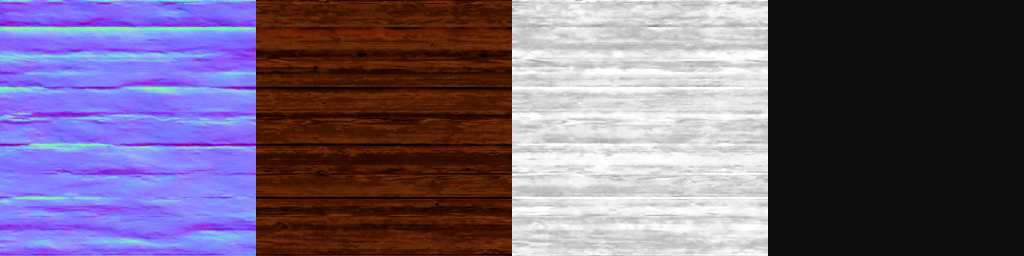}
            \includegraphics[width=\linewidth]{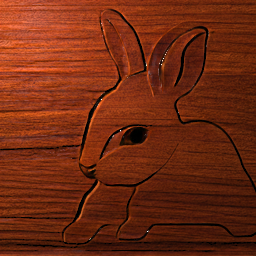} 
            \includegraphics[width=\linewidth]{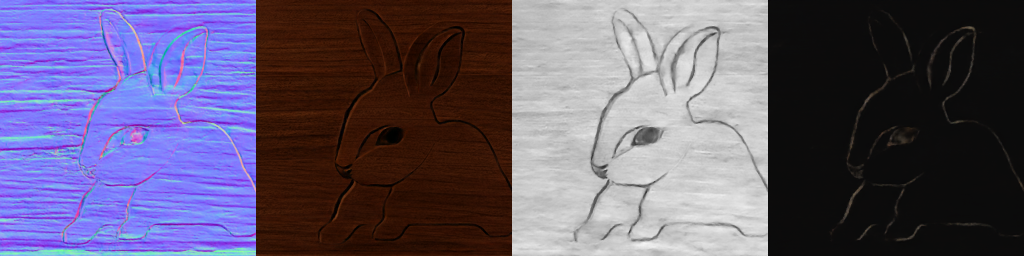}  
        \end{subfigure}
    \end{minipage}
    \caption{\textbf{Ablation Study.} Comparison of the current method with two prior approaches described in the Experiments section: (a) Training only the VAE decoder to output SVBRDF maps results in invalid maps despite understanding text beyond the training set. (b) A retrained single diffusion model yields slightly better-quality maps but fails to comprehend text outside the training set.}
    \label{fig:ablation}
    
\end{figure}

\begin{figure*}[h!]
    \centering
    \setlength{\tabcolsep}{2pt}
    
    \begin{tabular}{ccc}

    \begin{minipage}[h]{0.21\textwidth}
        \centering
        \includegraphics[width=\linewidth, trim={2cm 3cm 2cm 1cm},clip]{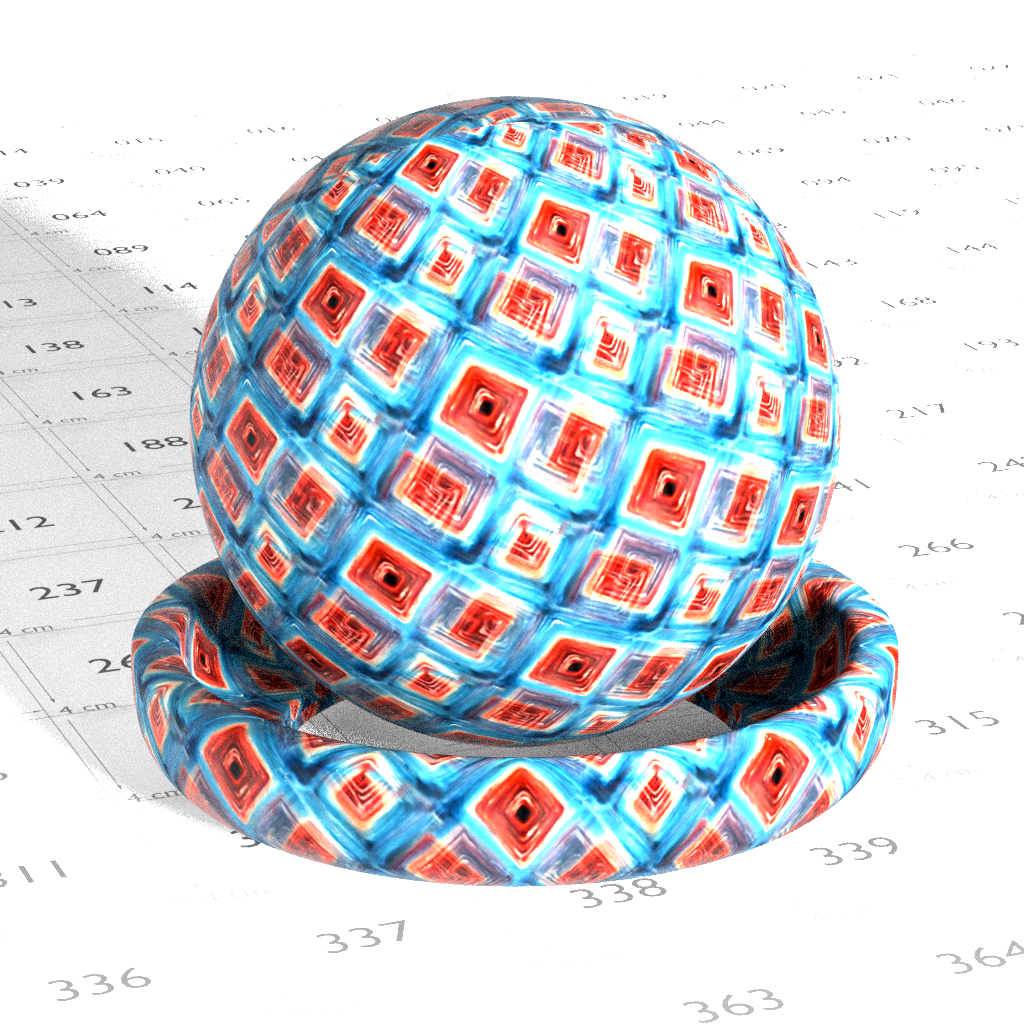}
        \input{image3/run_764/prompt.txt}
    \end{minipage}
    \begin{minipage}[h]{0.105\textwidth}
        \centering
        \includegraphics[width=\linewidth]{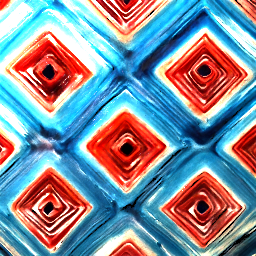}
        \includegraphics[width=\linewidth]{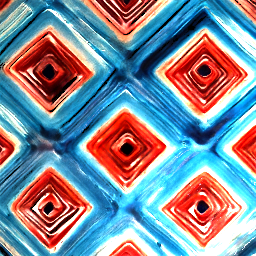}
        \includegraphics[width=\linewidth]{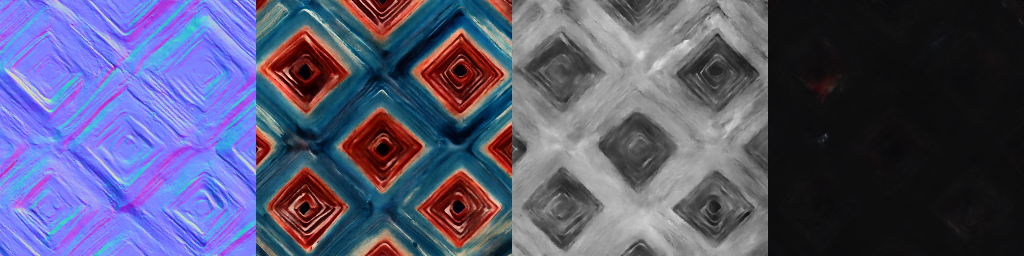}
        
        \hfill
    \end{minipage}

    \begin{minipage}[h]{0.21\textwidth}
        \centering
        \includegraphics[width=\linewidth, trim={2cm 3cm 2cm 1cm},clip]{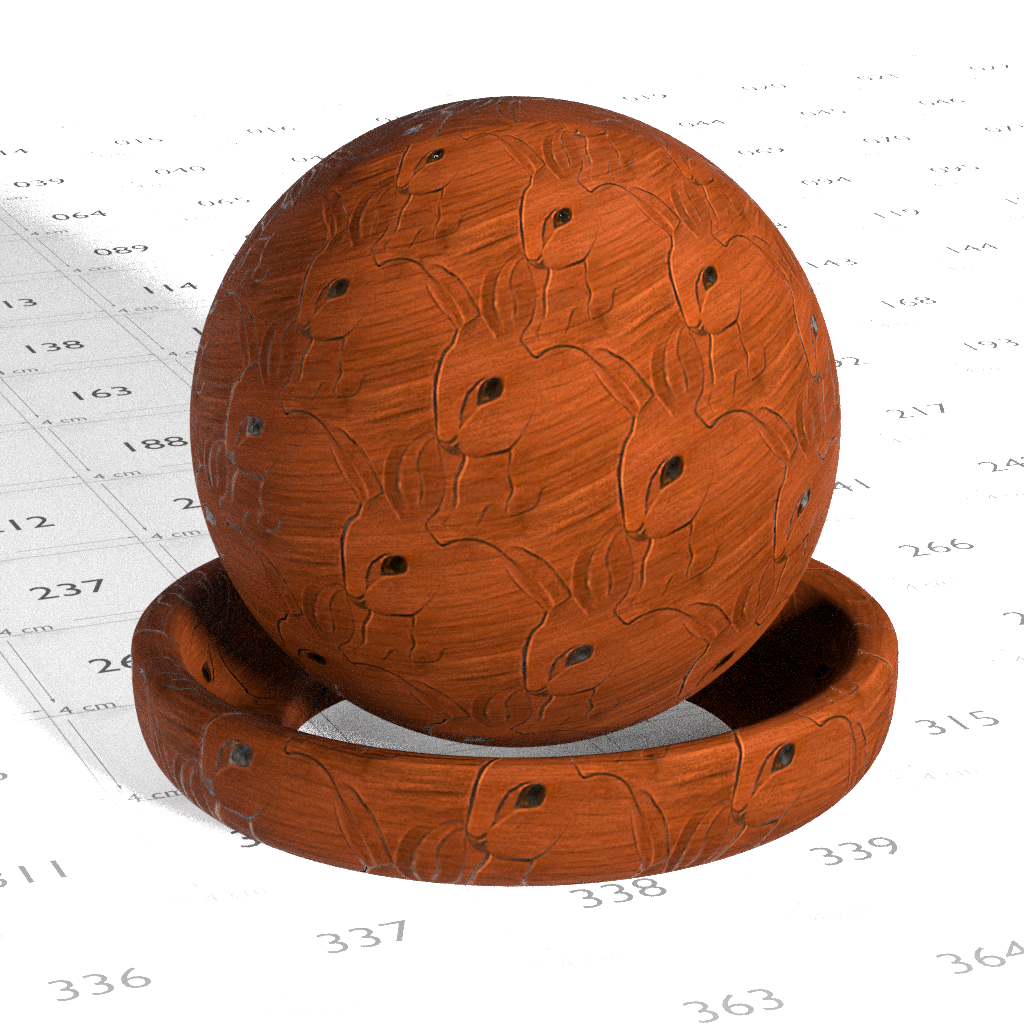}
        \input{image3/run_0441/prompt.txt}
    \end{minipage}
    \begin{minipage}[h]{0.105\textwidth}
        \centering
        \includegraphics[width=\linewidth]{image3/run_0441/00.png}
        \includegraphics[width=\linewidth]{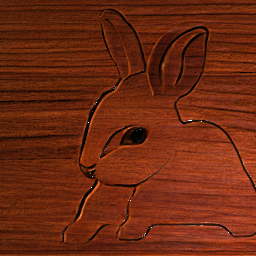}
        \includegraphics[width=\linewidth]{image3/run_0441/renderimage1d.png}
        
        \hfill
    \end{minipage}

    \begin{minipage}[h]{0.21\textwidth}
        \centering
        \includegraphics[width=\linewidth, trim={2cm 3cm 2cm 1cm},clip]{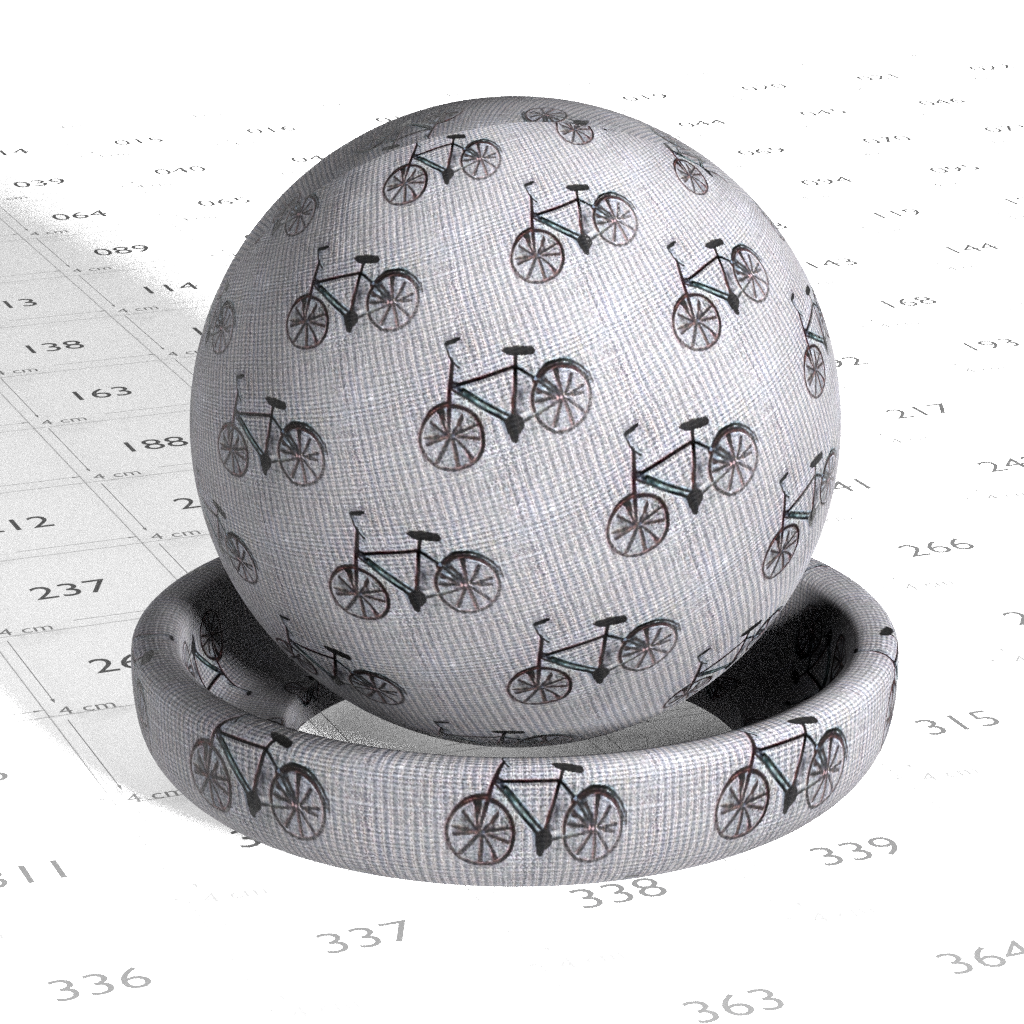}
        \input{image3/run_804/prompt.txt}
    \end{minipage}
    \begin{minipage}[h]{0.105\textwidth}
        \centering
        \includegraphics[width=\linewidth]{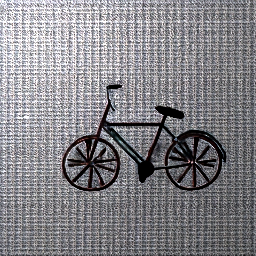}
        \includegraphics[width=\linewidth]{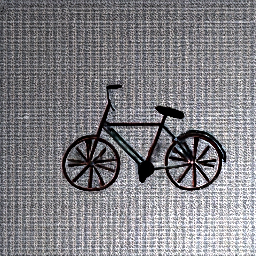}
        \includegraphics[width=\linewidth]{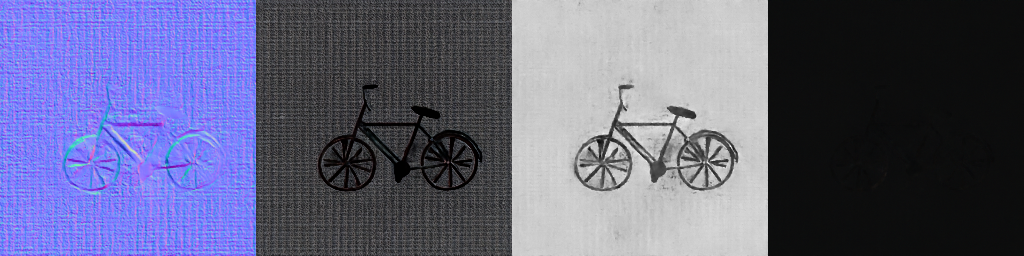}
        
        \hfill
    \end{minipage}

    \begin{minipage}[h]{0.21\textwidth}
        \centering
        \includegraphics[width=\linewidth, trim={2cm 3cm 2cm 1cm},clip]{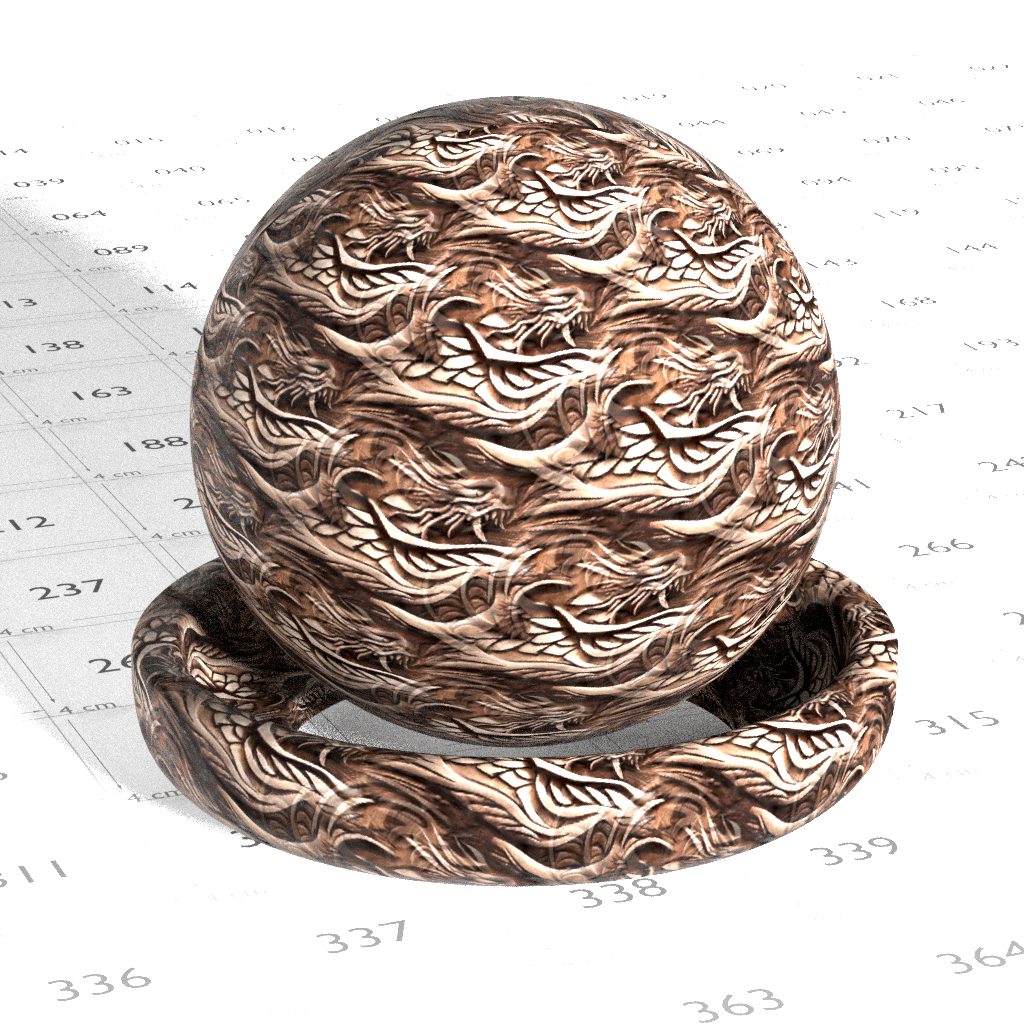}
        \input{image3/run_601/prompt.txt}
    \end{minipage}
    \begin{minipage}[h]{0.105\textwidth}
        \centering
        \includegraphics[width=\linewidth]{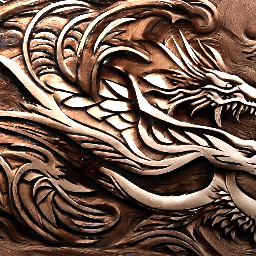}
        \includegraphics[width=\linewidth]{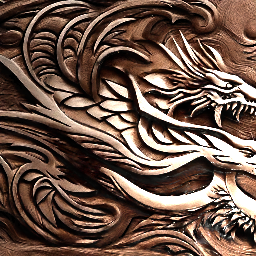}
        \includegraphics[width=\linewidth]{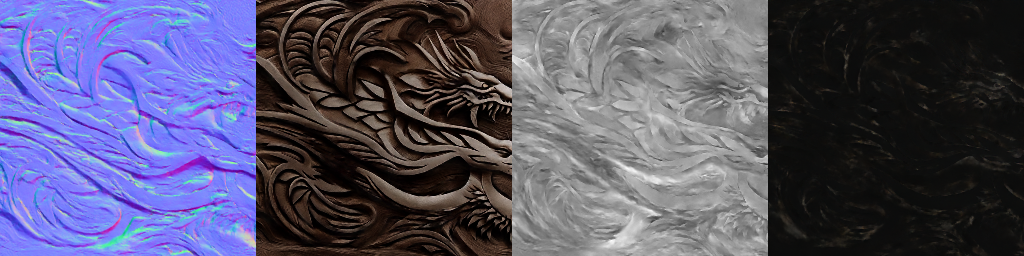}
        
        \hfill
    \end{minipage}

    \begin{minipage}[h]{0.21\textwidth}
        \centering
        \includegraphics[width=\linewidth, trim={2cm 3cm 2cm 1cm},clip]{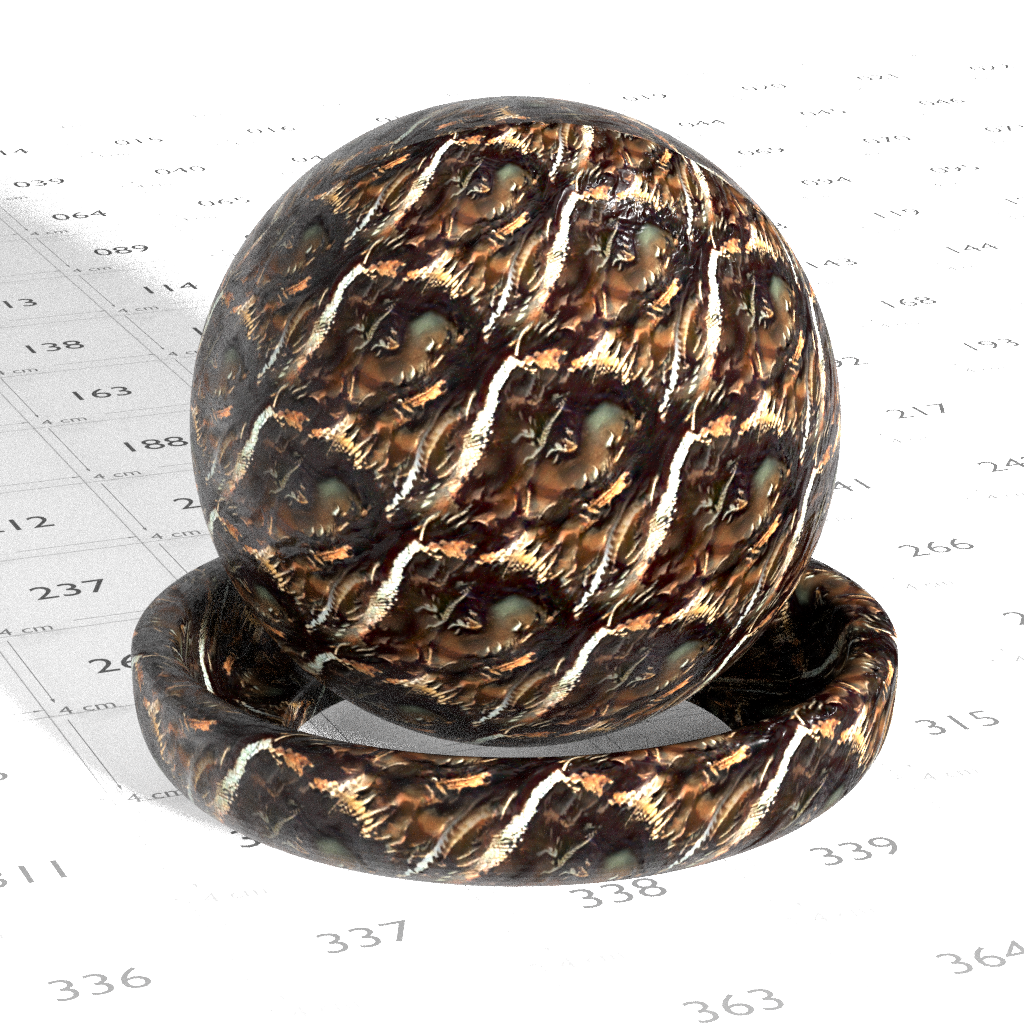}
        \input{image3/run_754/prompt.txt}
    \end{minipage}
    \begin{minipage}[h]{0.105\textwidth}
        \centering
        \includegraphics[width=\linewidth]{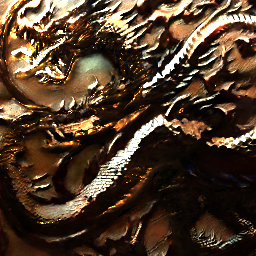}
        \includegraphics[width=\linewidth]{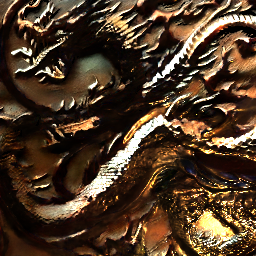}
        \includegraphics[width=\linewidth]{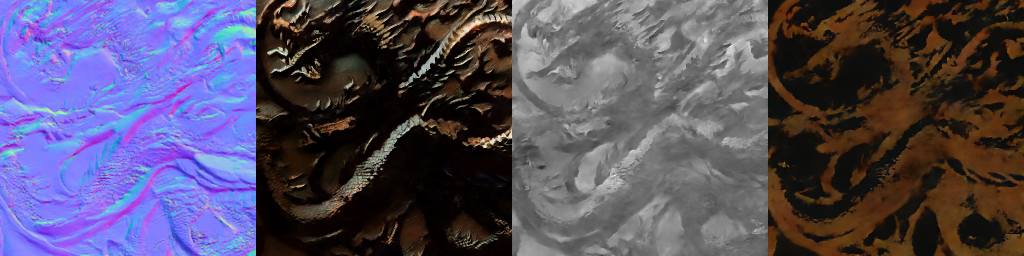}
        
        \hfill
    \end{minipage}

    \begin{minipage}[h]{0.21\textwidth}
        \centering
        \includegraphics[width=\linewidth, trim={2cm 3cm 2cm 1cm},clip]{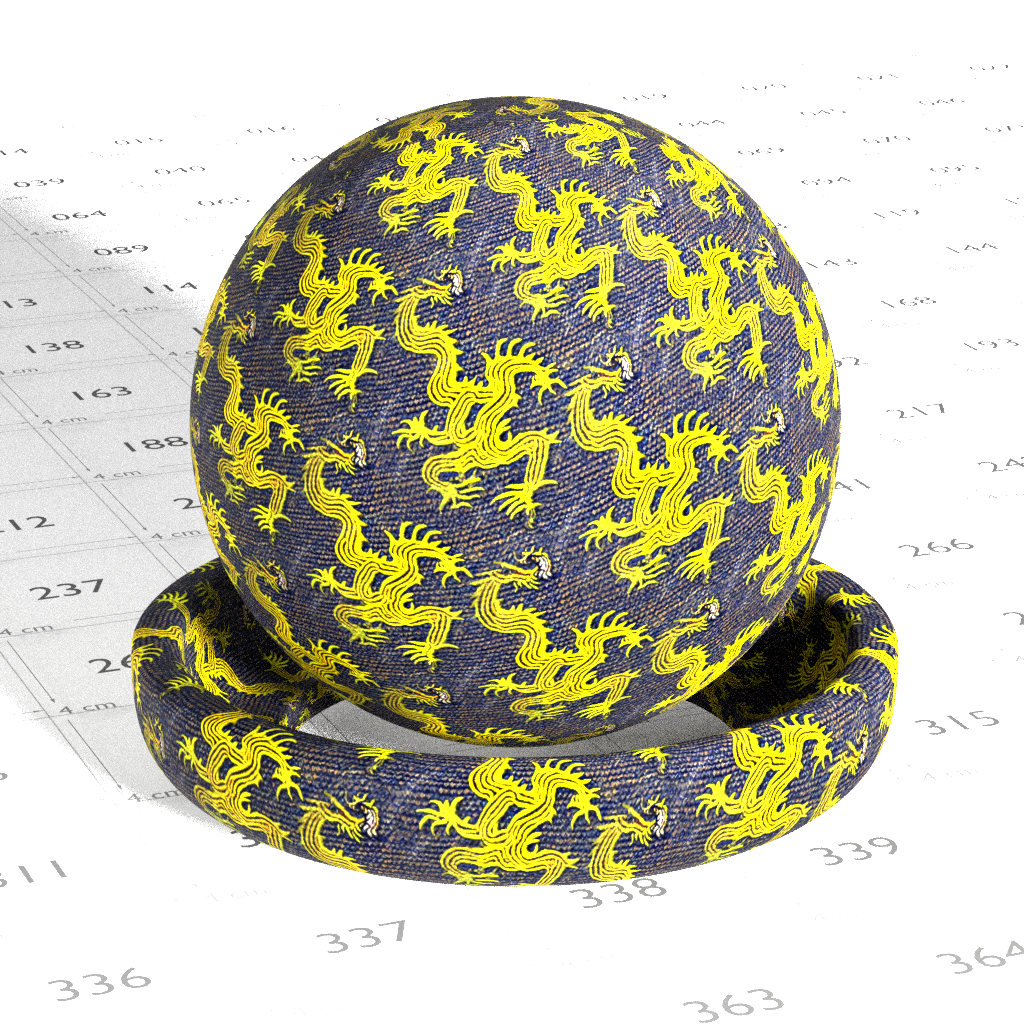}
        \input{image3/run_839/prompt.txt}
    \end{minipage}
    \begin{minipage}[h]{0.105\textwidth}
        \centering
        \includegraphics[width=\linewidth]{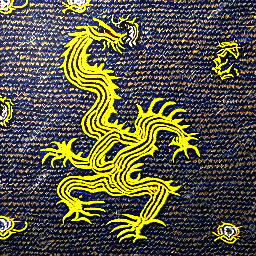}
        \includegraphics[width=\linewidth]{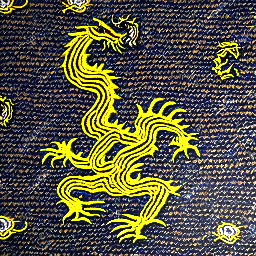}
        \includegraphics[width=\linewidth]{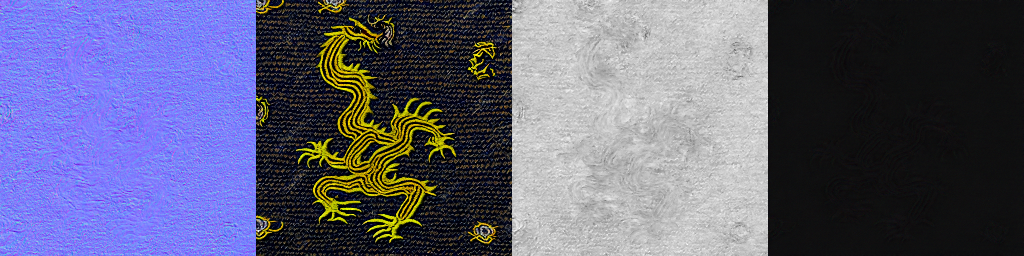}
        
        \hfill
    \end{minipage}

    \begin{minipage}[h]{0.21\textwidth}
        \centering
        \includegraphics[width=\linewidth, trim={2cm 3cm 2cm 1cm},clip]{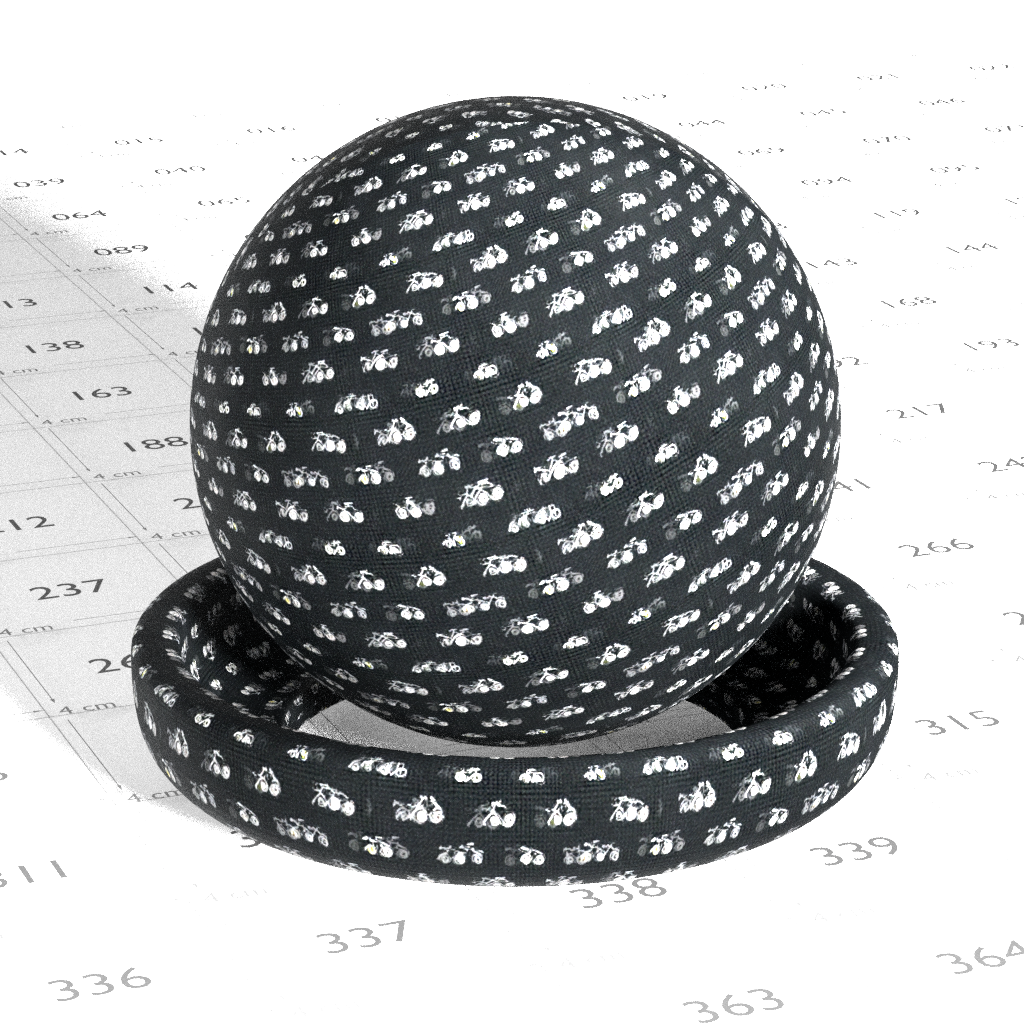}
        \input{image3/run_811/prompt.txt}
    \end{minipage}
    \begin{minipage}[h]{0.105\textwidth}
        \centering
        \includegraphics[width=\linewidth]{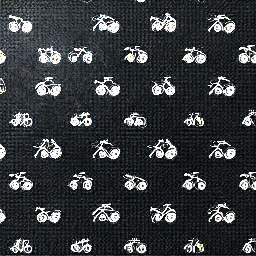}
        \includegraphics[width=\linewidth]{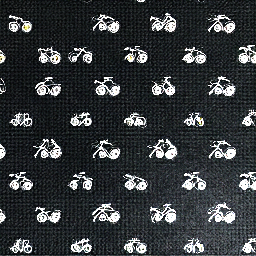}
        \includegraphics[width=\linewidth]{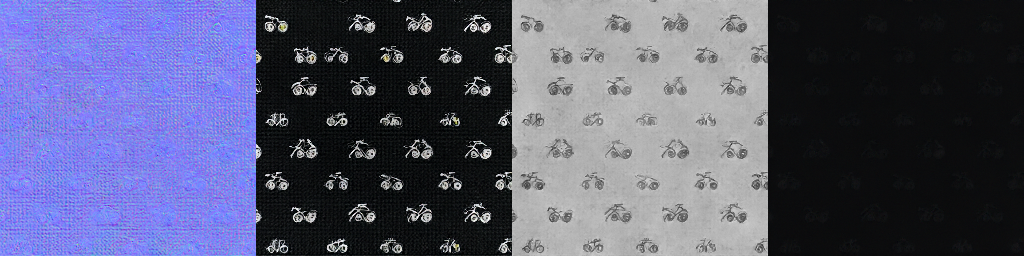}
        
        \hfill
    \end{minipage}

    \begin{minipage}[h]{0.21\textwidth}
        \centering
        \includegraphics[width=\linewidth, trim={2cm 3cm 2cm 1cm},clip]{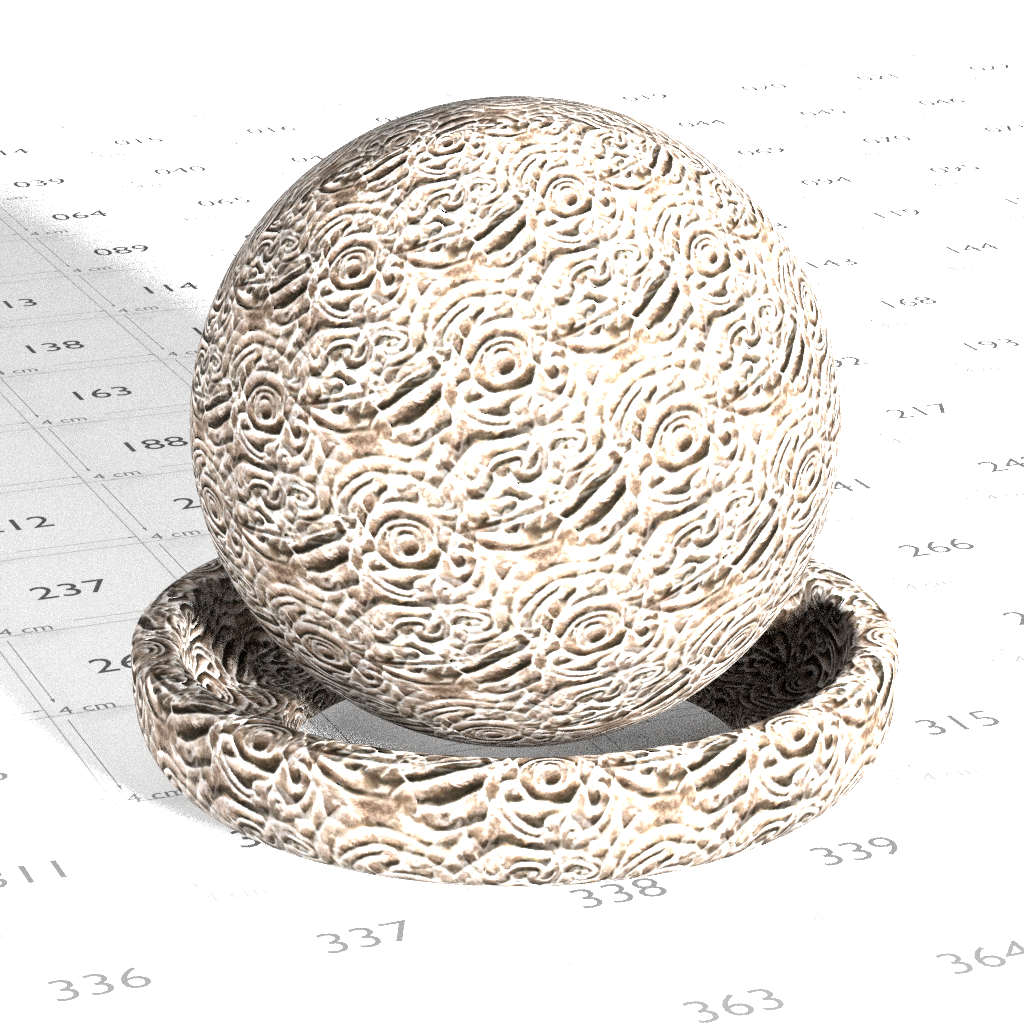}
        \input{image3/run_251/prompt.txt}
    \end{minipage}
    \begin{minipage}[h]{0.105\textwidth}
        \centering
        \includegraphics[width=\linewidth]{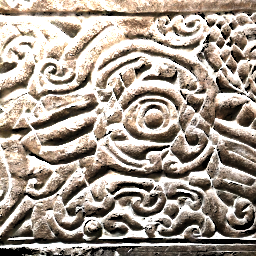}
        \includegraphics[width=\linewidth]{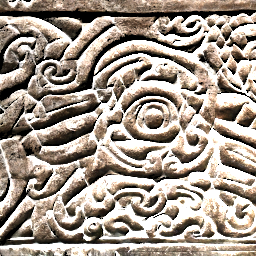}
        \includegraphics[width=\linewidth]{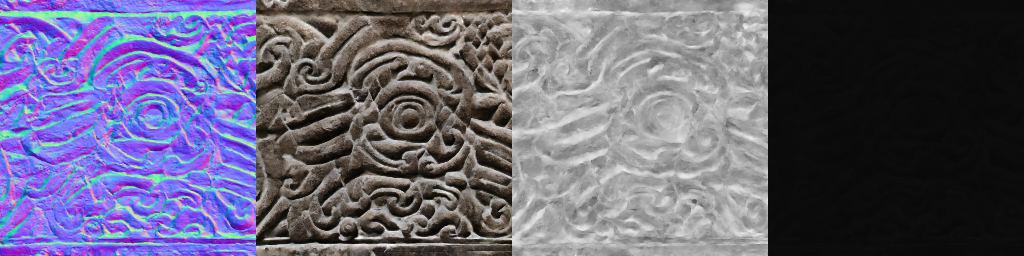}
        
        \hfill
    \end{minipage}

    \begin{minipage}[h]{0.21\textwidth}
        \centering
        \includegraphics[width=\linewidth, trim={2cm 3cm 2cm 1cm},clip]{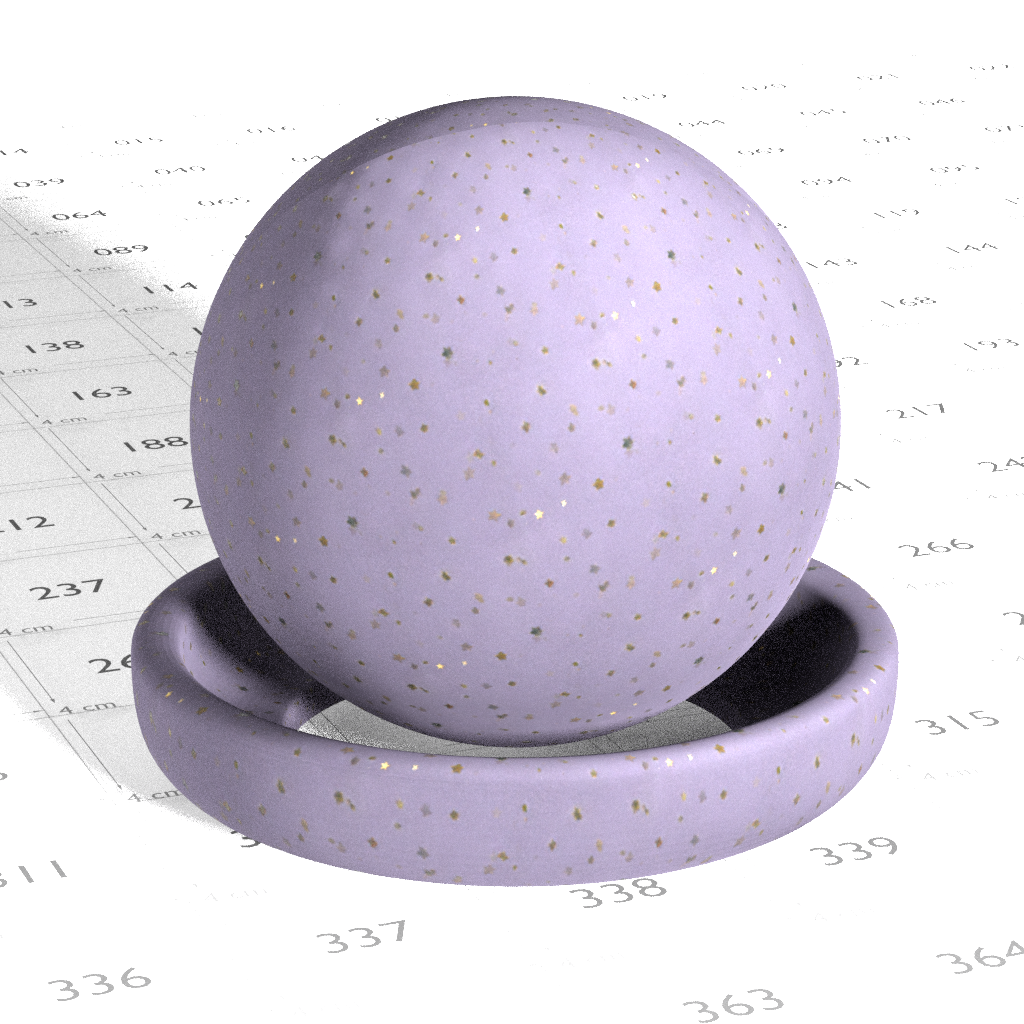}
        \input{image3/run_211/prompt.txt}
    \end{minipage}
    \begin{minipage}[h]{0.105\textwidth}
        \centering
        \includegraphics[width=\linewidth]{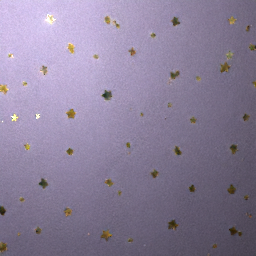}
        \includegraphics[width=\linewidth]{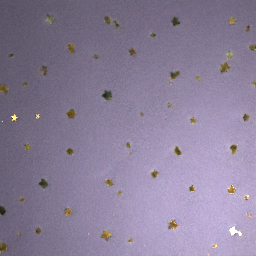}
        \includegraphics[width=\linewidth]{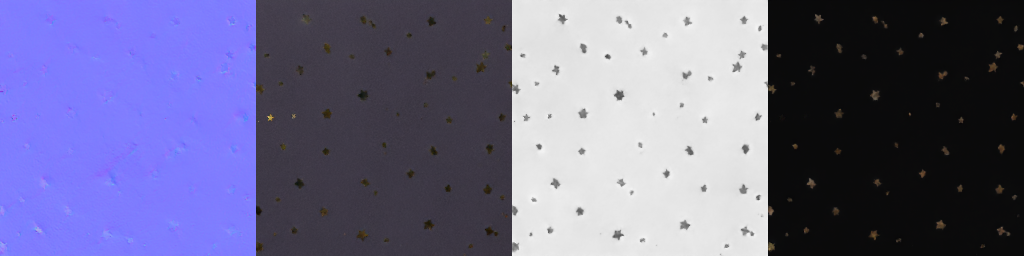}
        
        \hfill
    \end{minipage}

    \begin{minipage}[h]{0.21\textwidth}
        \centering
        \includegraphics[width=\linewidth, trim={2cm 3cm 2cm 1cm},clip]{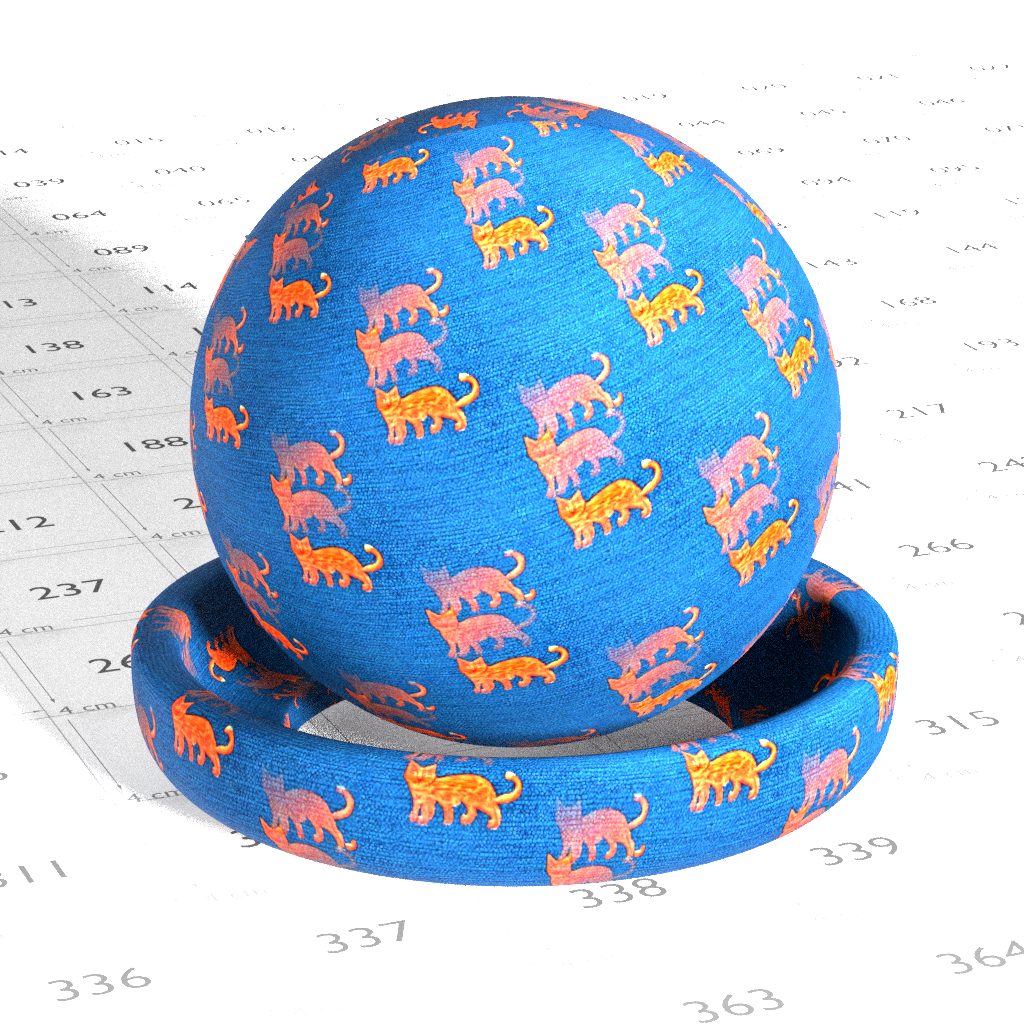}
        \input{image3/run_819/prompt.txt}
    \end{minipage}
    \begin{minipage}[h]{0.105\textwidth}
        \centering
        \includegraphics[width=\linewidth]{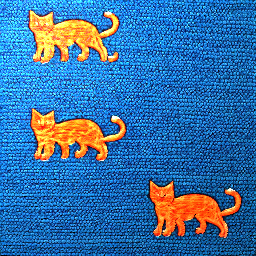}
        \includegraphics[width=\linewidth]{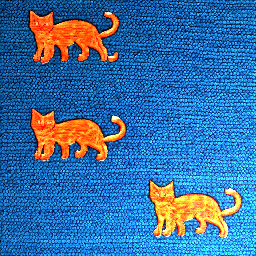}
        \includegraphics[width=\linewidth]{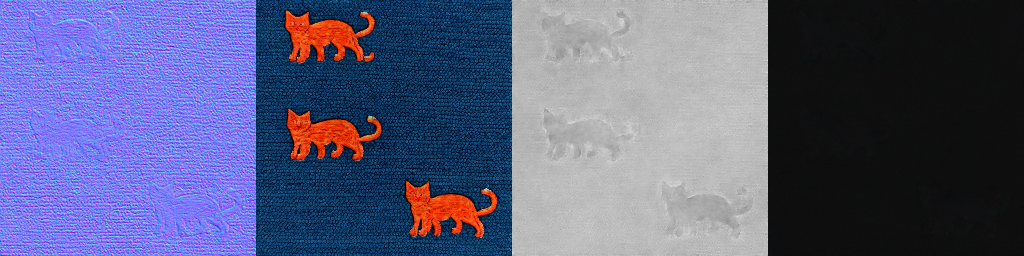}
        
        \hfill
    \end{minipage}

    \begin{minipage}[h]{0.21\textwidth}
        \centering
        \includegraphics[width=\linewidth, trim={2cm 3cm 2cm 1cm},clip]{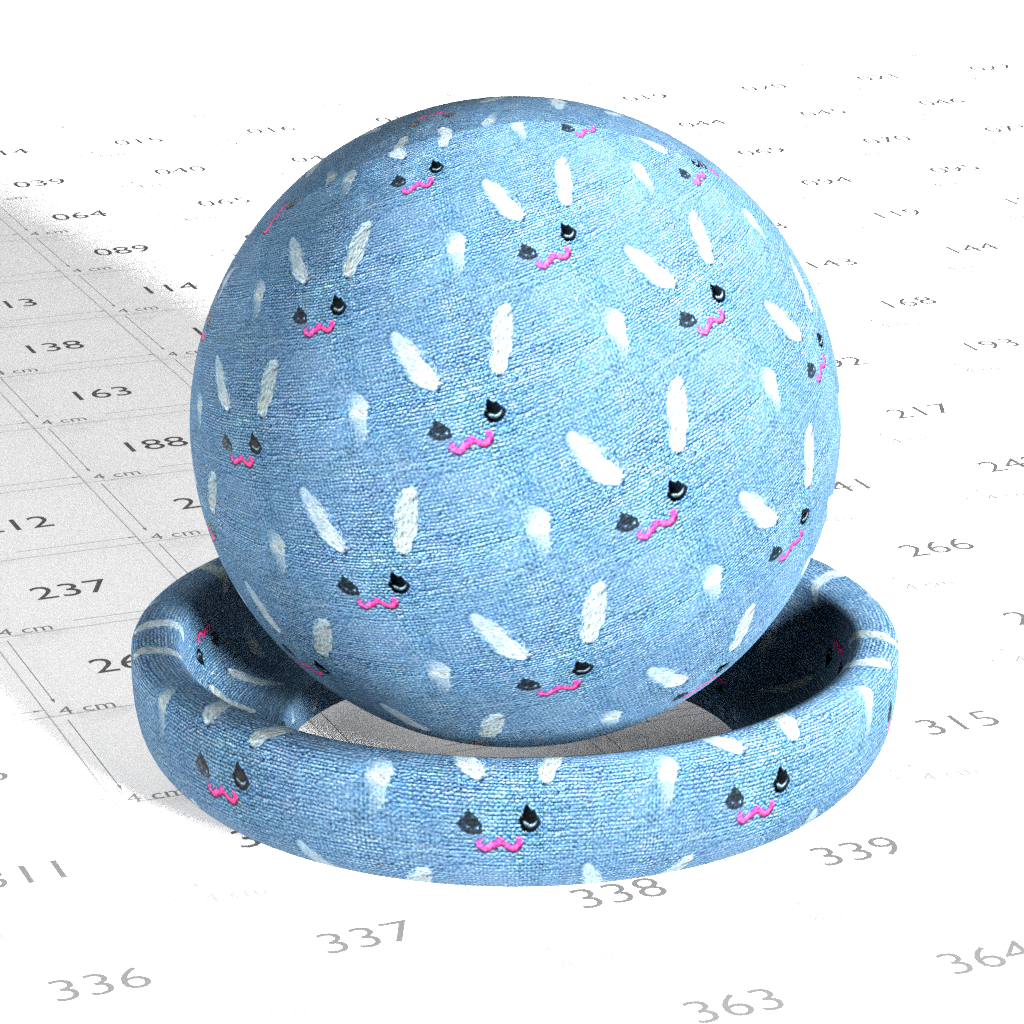}
        \input{image3/run_588/prompt.txt}
    \end{minipage}
    \begin{minipage}[h]{0.105\textwidth}
        \centering
        \includegraphics[width=\linewidth]{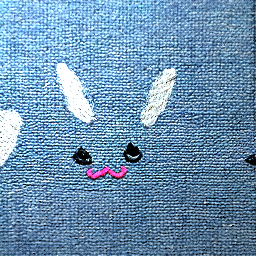}
        \includegraphics[width=\linewidth]{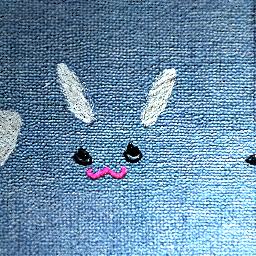}
        \includegraphics[width=\linewidth]{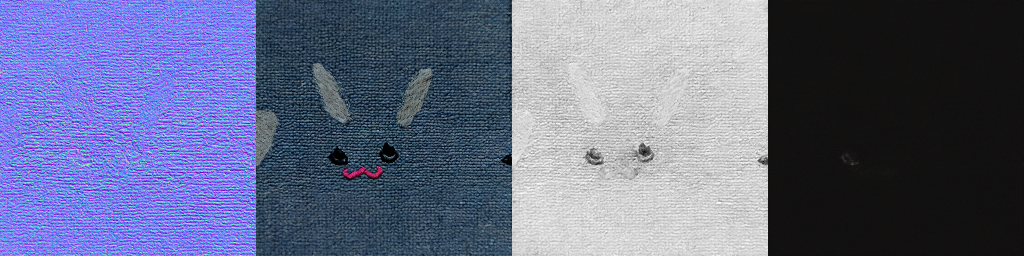}
        
        \hfill
    \end{minipage}

    \begin{minipage}[h]{0.21\textwidth}
        \centering
        \includegraphics[width=\linewidth, trim={2cm 3cm 2cm 1cm},clip]{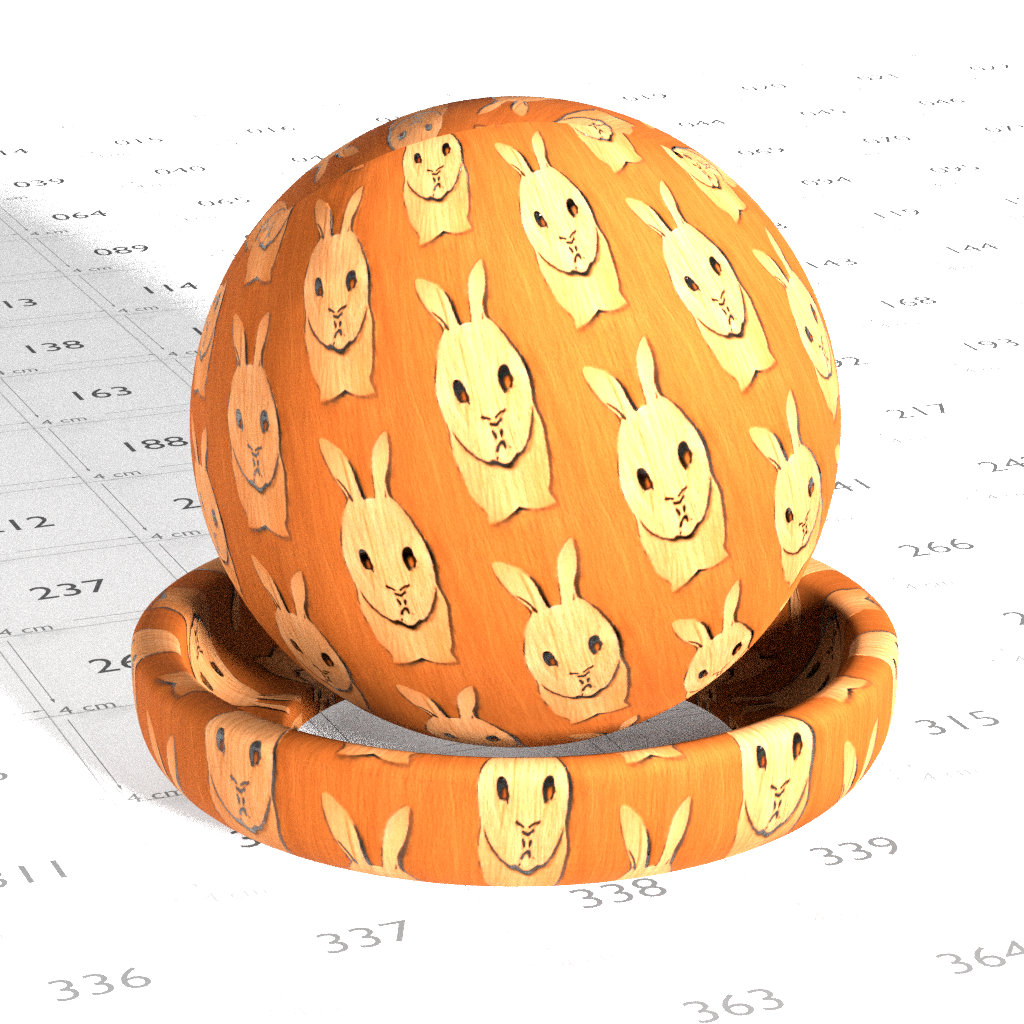}
        \input{image3/run_0730/prompt.txt}
    \end{minipage}
    \begin{minipage}[h]{0.105\textwidth}
        \centering
        \includegraphics[width=\linewidth]{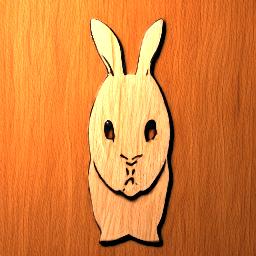}
        \includegraphics[width=\linewidth]{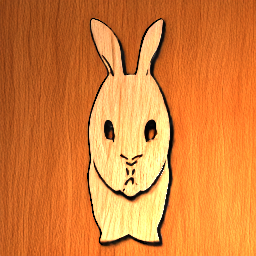}
        \includegraphics[width=\linewidth]{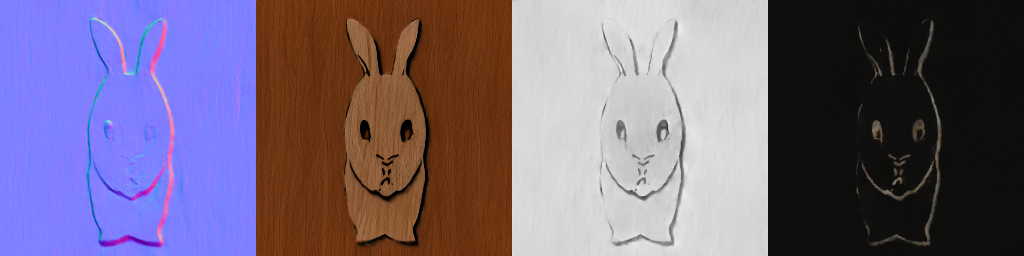}
        
        \hfill
    \end{minipage}

    \end{tabular}
    \caption{\textbf{Non-stationary results.} We demonstrate our method for generating SVBRDF maps, with two flat renderings illuminated from the top left and bottom right corners shown on the right. The same texture, tiled and applied to a 3D shape, is displayed on the left.}
    \label{fig_mainresult} 
\end{figure*}
\begin{figure*}[h!]
    \centering
    \setlength{\tabcolsep}{2pt}
    
    \begin{tabular}{cc}

    \begin{minipage}[h]{0.21\textwidth}
        \centering
        \includegraphics[width=\linewidth, trim={2cm 3cm 2cm 1cm},clip]{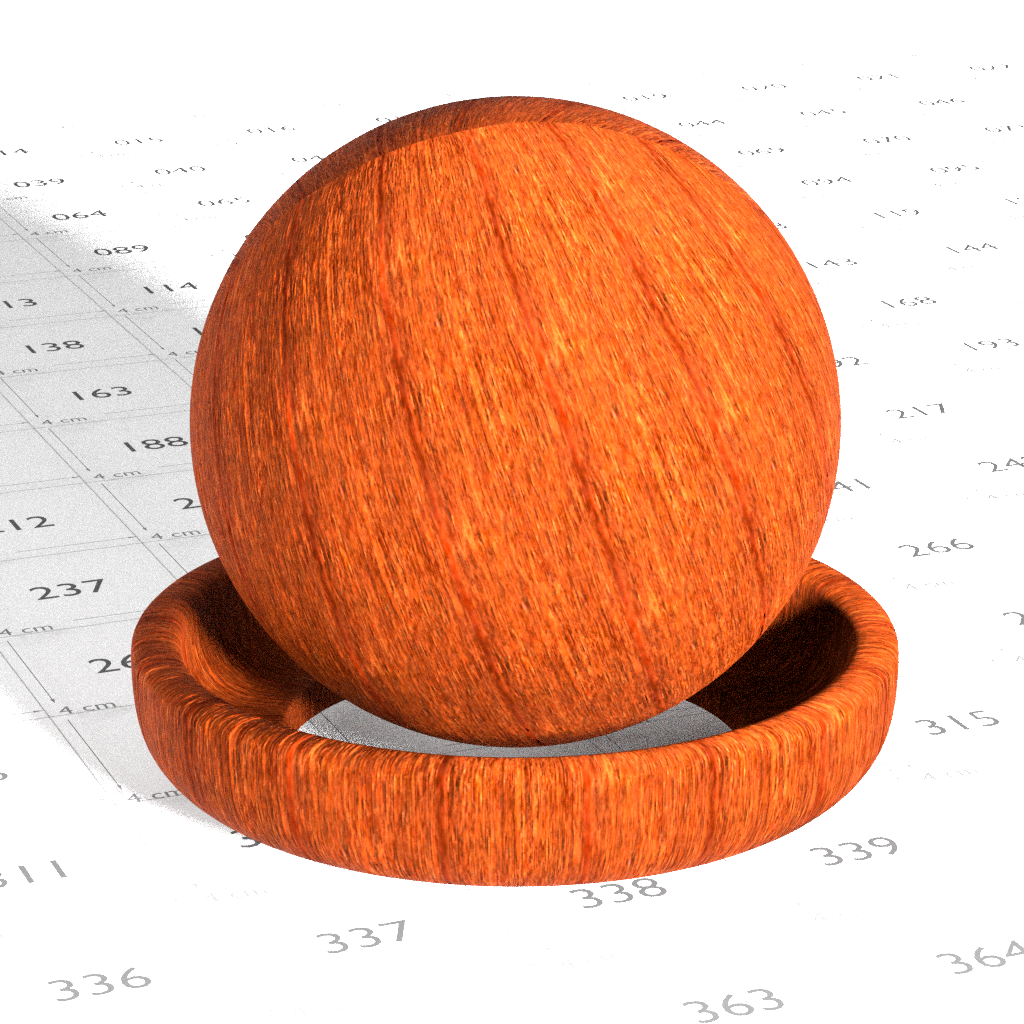}
        \input{image3/run_225/prompt.txt}
    \end{minipage}
    \begin{minipage}[h]{0.105\textwidth}
        \centering
        \includegraphics[width=\linewidth]{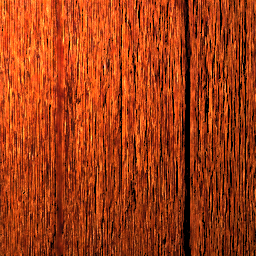}
        \includegraphics[width=\linewidth]{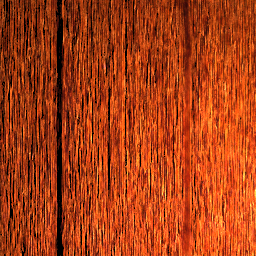}
        \includegraphics[width=\linewidth]{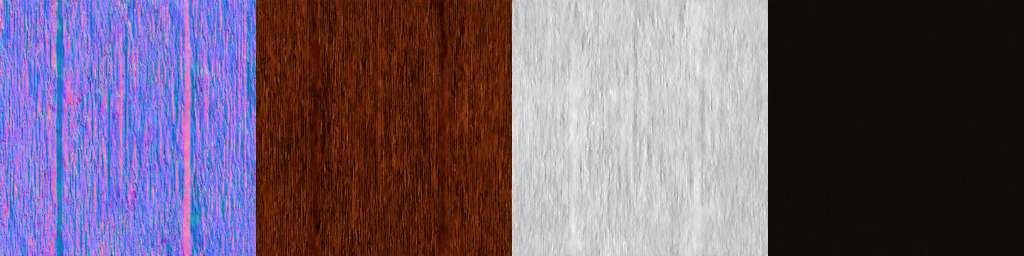}
        
        \hfill
    \end{minipage}

    \begin{minipage}[h]{0.21\textwidth}
        \centering
        \includegraphics[width=\linewidth, trim={2cm 3cm 2cm 1cm},clip]{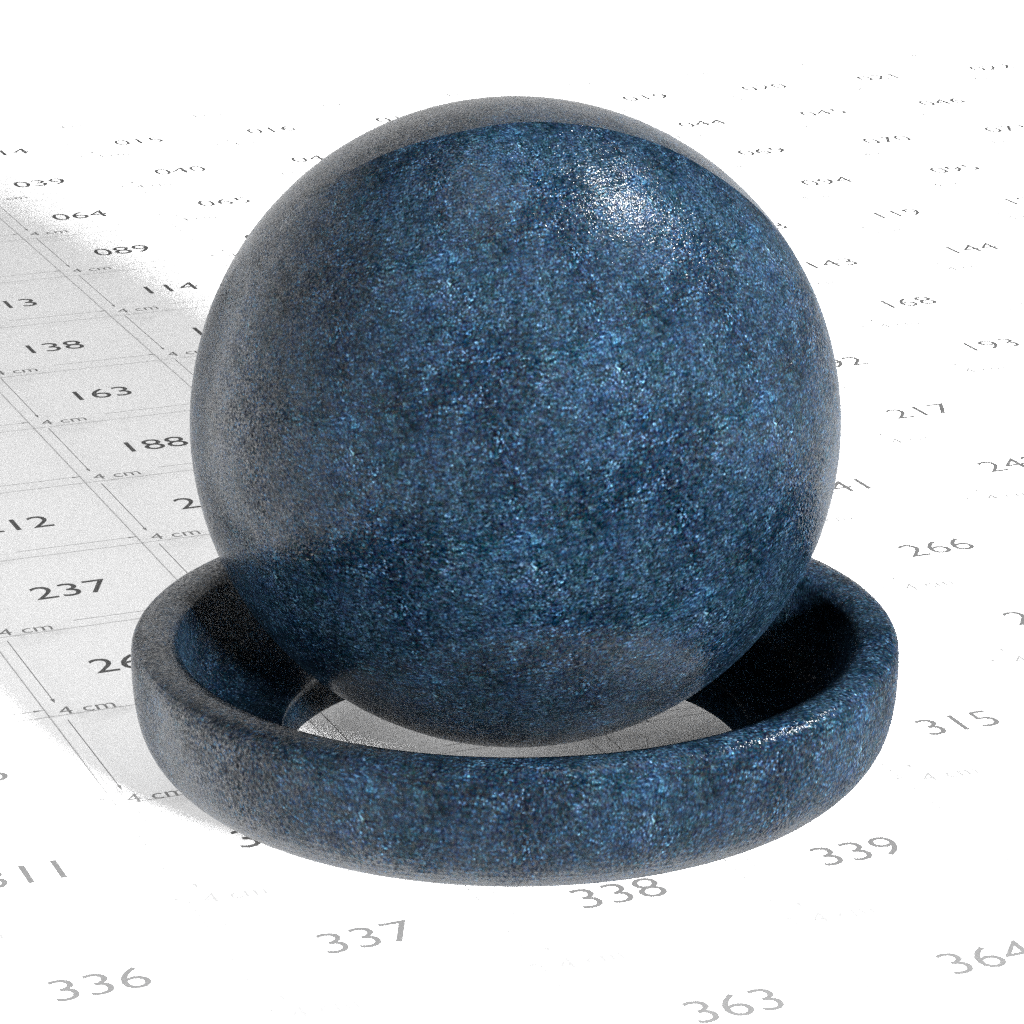}
        \input{image3/run_576/prompt.txt}
    \end{minipage}
    \begin{minipage}[h]{0.105\textwidth}
        \centering
        \includegraphics[width=\linewidth]{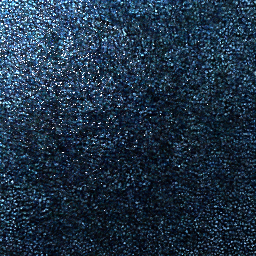}
        \includegraphics[width=\linewidth]{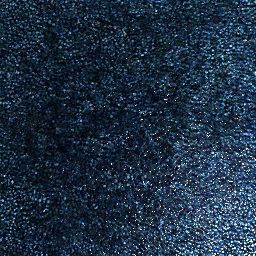}
        \includegraphics[width=\linewidth]{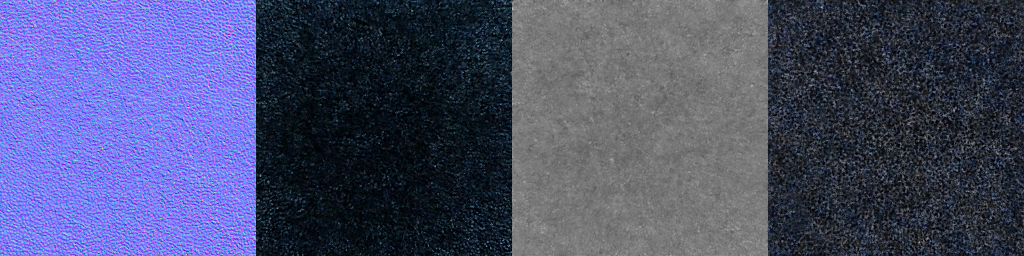}
        
        \hfill
    \end{minipage}

    \begin{minipage}[h]{0.21\textwidth}
        \centering
        \includegraphics[width=\linewidth, trim={2cm 3cm 2cm 1cm},clip]{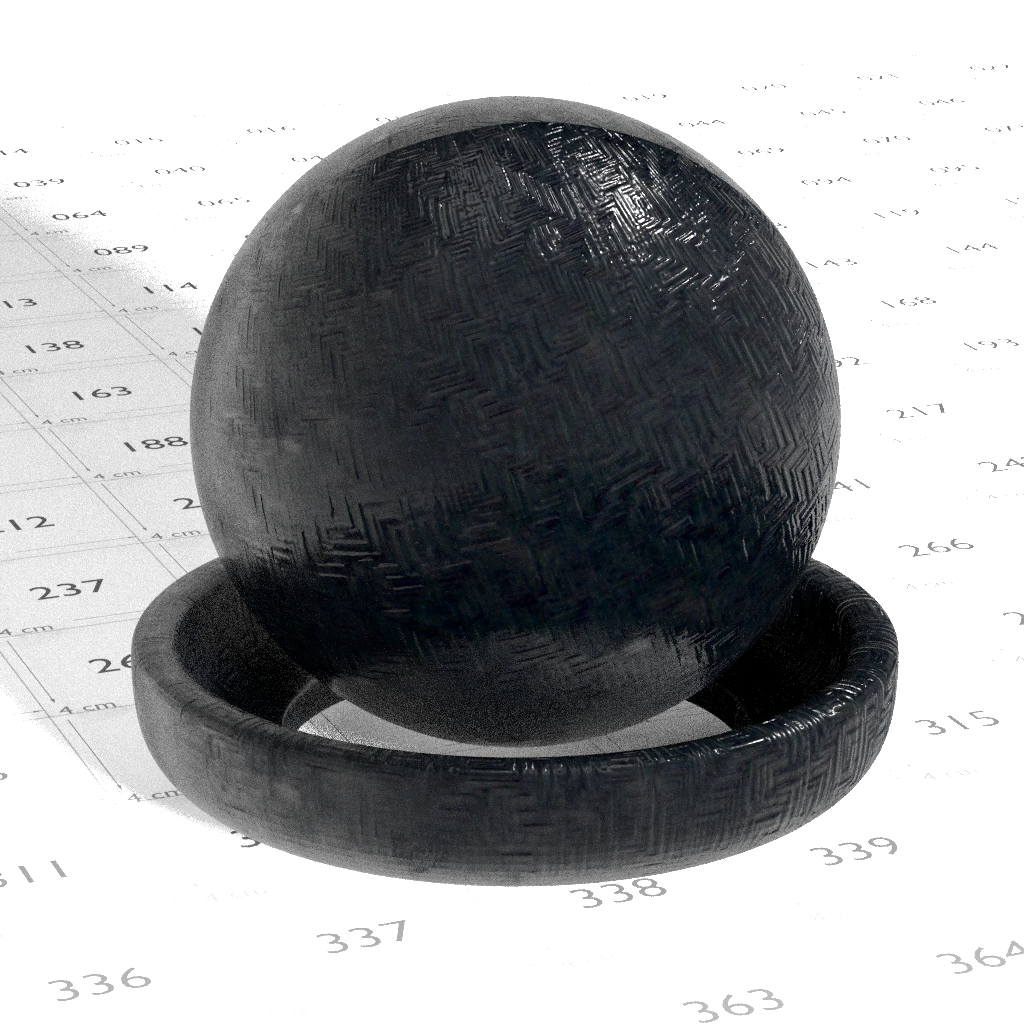}
        \input{image3/run_240/prompt.txt}
    \end{minipage}
    \begin{minipage}[h]{0.105\textwidth}
        \centering
        \includegraphics[width=\linewidth]{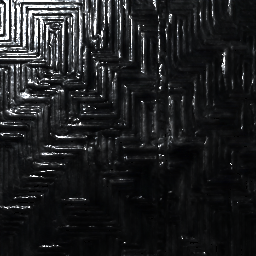}
        \includegraphics[width=\linewidth]{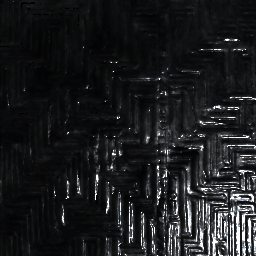}
        \includegraphics[width=\linewidth]{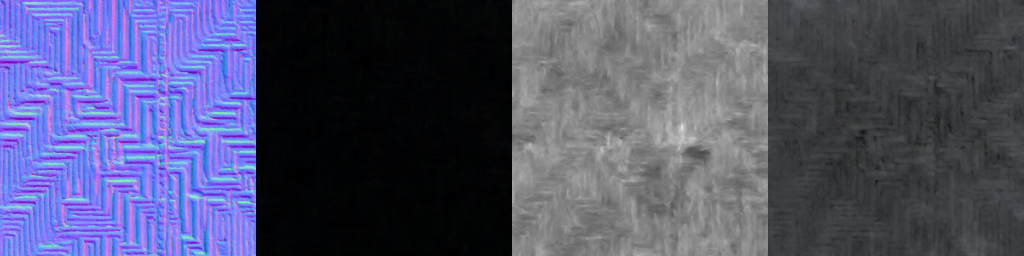}
        
        \hfill
    \end{minipage}

    \begin{minipage}[h]{0.21\textwidth}
        \centering
        \includegraphics[width=\linewidth, trim={2cm 3cm 2cm 1cm},clip]{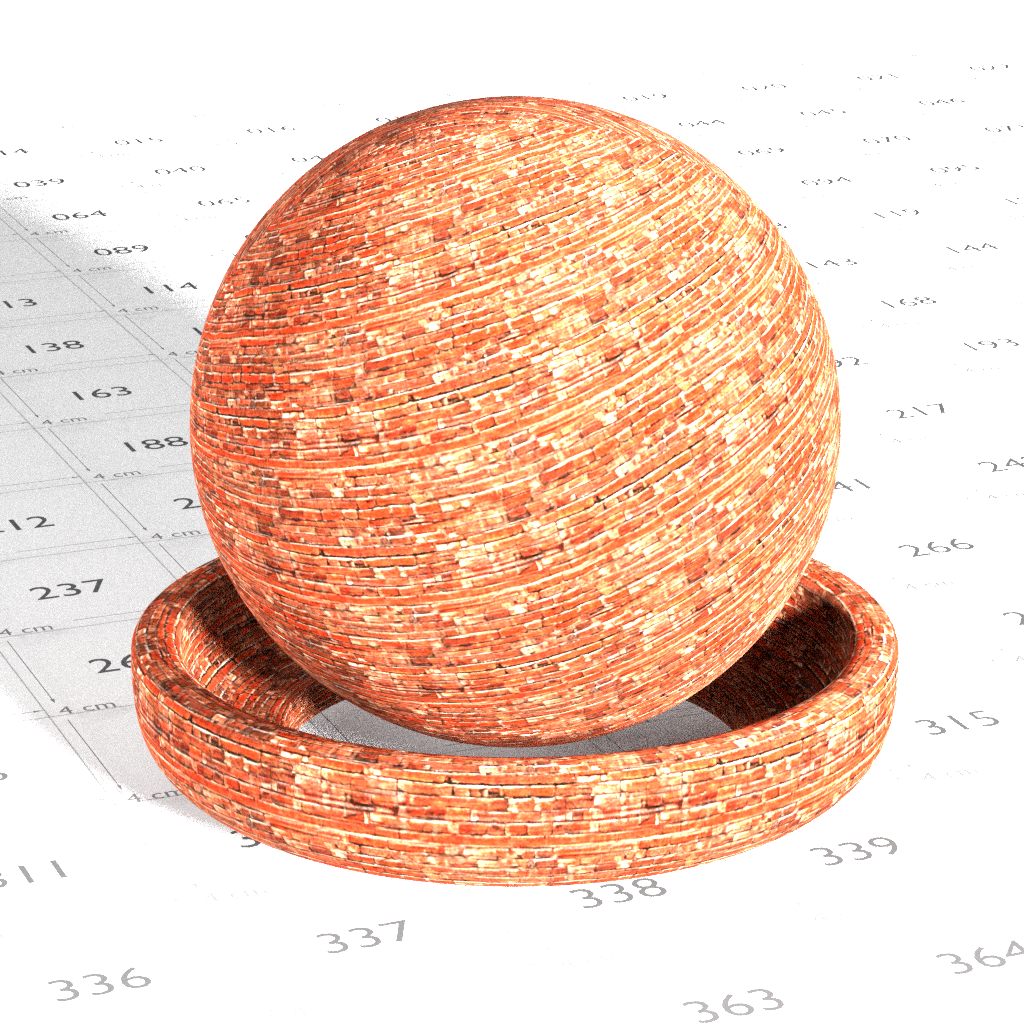}
        \input{image3/run_0646/prompt.txt}
    \end{minipage}
    \begin{minipage}[h]{0.105\textwidth}
        \centering
        \includegraphics[width=\linewidth]{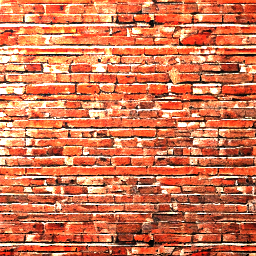}
        \includegraphics[width=\linewidth]{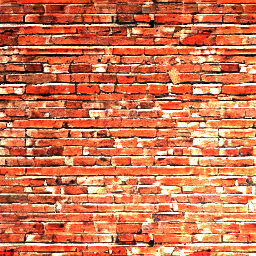}
        \includegraphics[width=\linewidth]{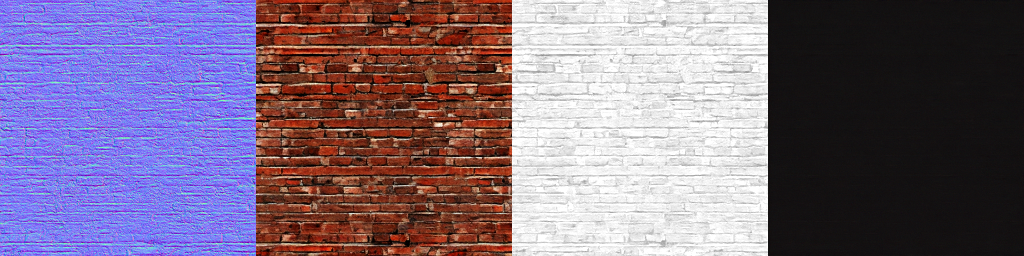}
        
        \hfill
    \end{minipage}

    \begin{minipage}[h]{0.21\textwidth}
        \centering
        \includegraphics[width=\linewidth, trim={2cm 3cm 2cm 1cm},clip]{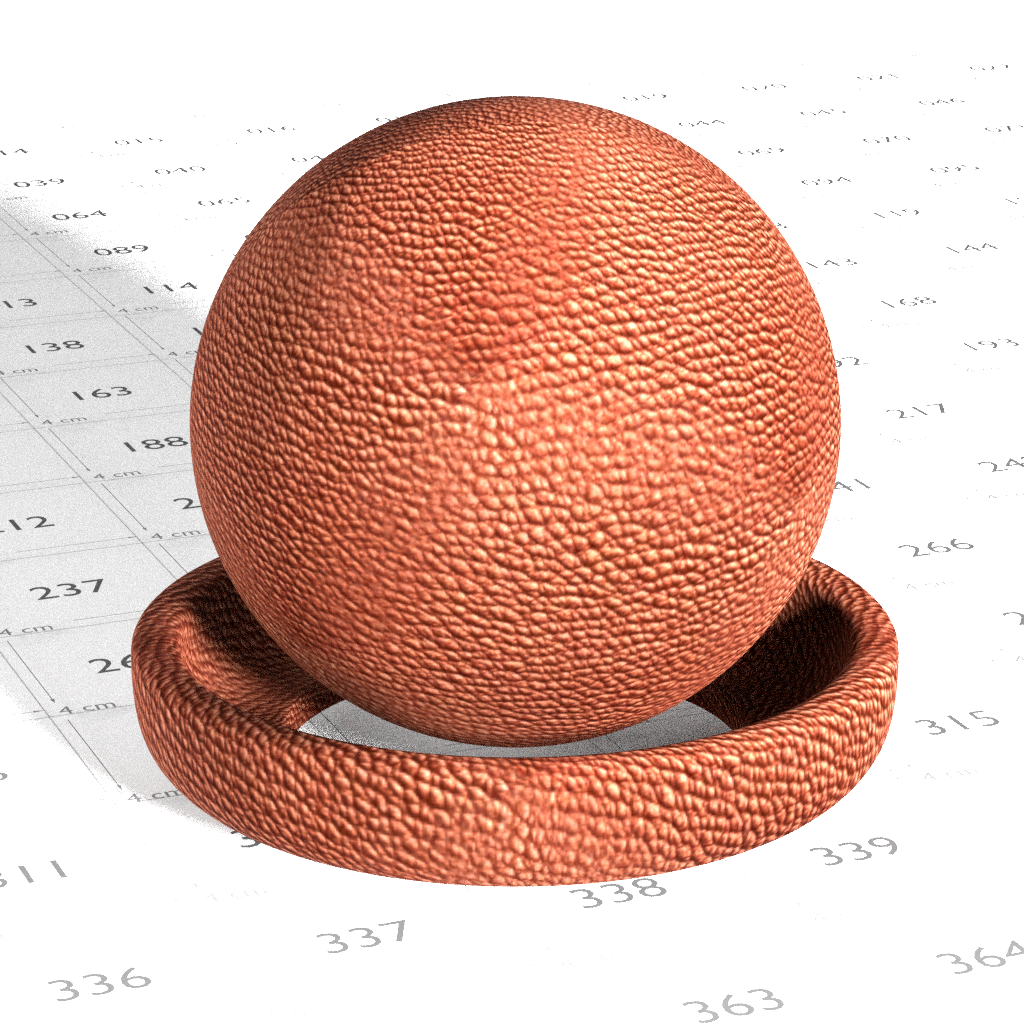}
        \input{image3/run_1328/prompt.txt}
    \end{minipage}
    \begin{minipage}[h]{0.105\textwidth}
        \centering
        \includegraphics[width=\linewidth]{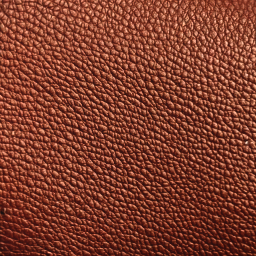}
        \includegraphics[width=\linewidth]{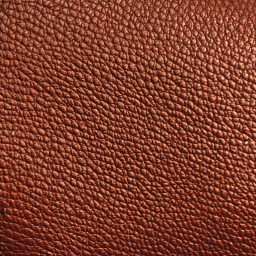}
        \includegraphics[width=\linewidth]{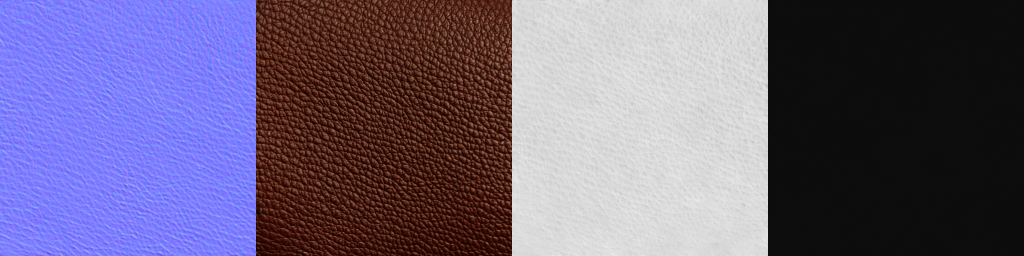}
        
        \hfill
    \end{minipage}

    \begin{minipage}[h]{0.21\textwidth}
        \centering
        \includegraphics[width=\linewidth, trim={2cm 3cm 2cm 1cm},clip]{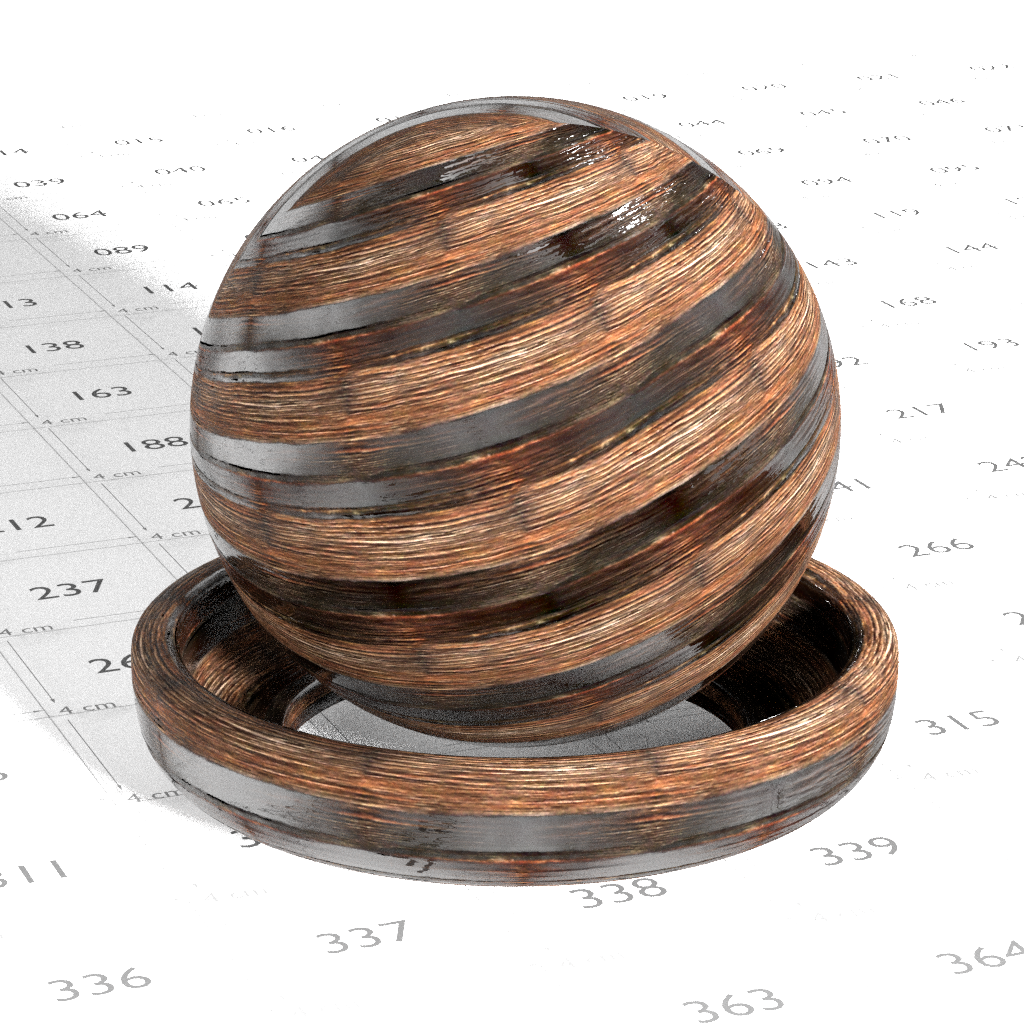}
        \input{image3/run_0426/prompt.txt}
    \end{minipage}
    \begin{minipage}[h]{0.105\textwidth}
        \centering
        \includegraphics[width=\linewidth]{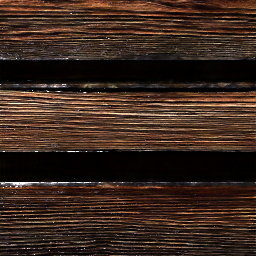}
        \includegraphics[width=\linewidth]{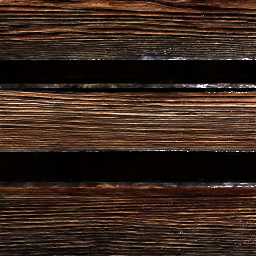}
        \includegraphics[width=\linewidth]{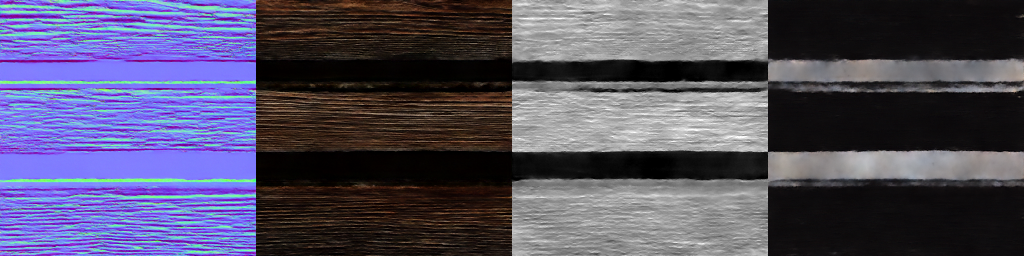}
        
        \hfill
    \end{minipage}

    \begin{minipage}[h]{0.21\textwidth}
        \centering
        \includegraphics[width=\linewidth, trim={2cm 3cm 2cm 1cm},clip]{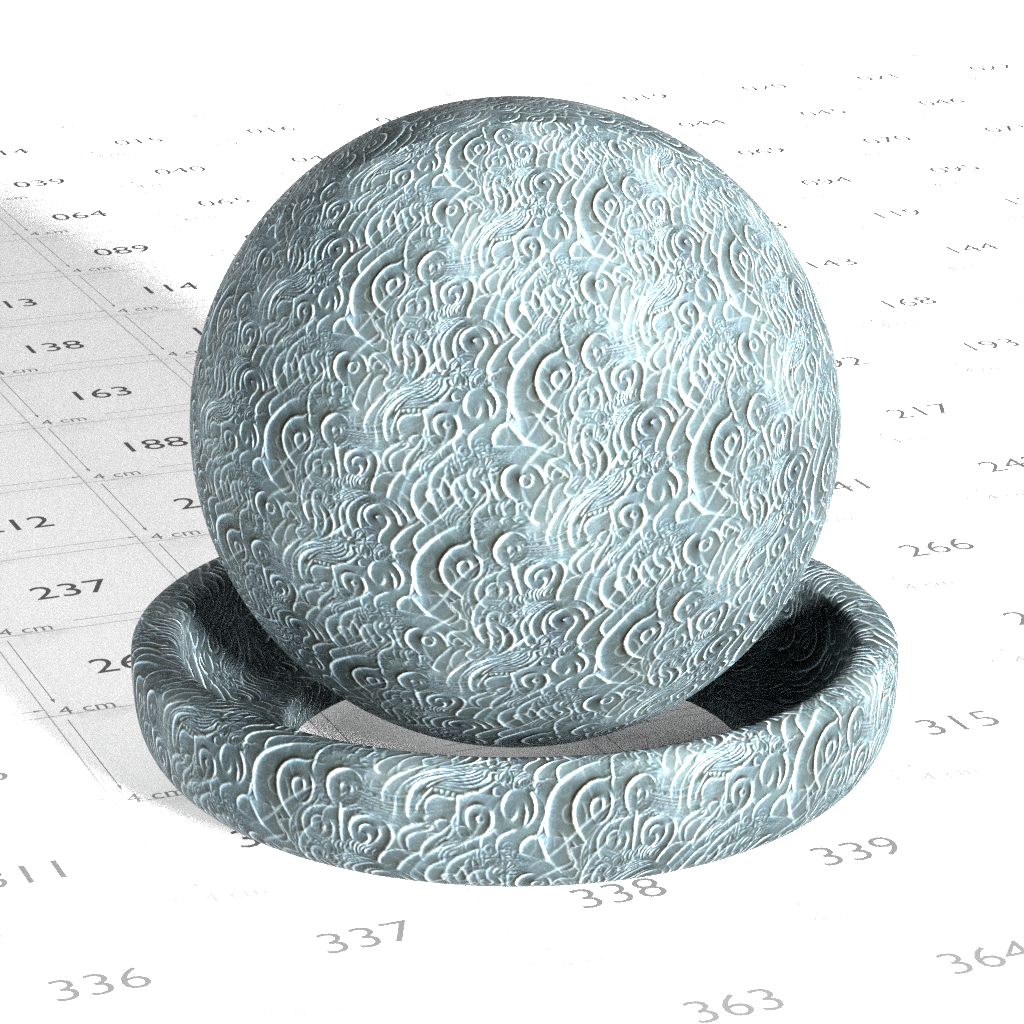}
        \input{image3/run_261/prompt.txt}
    \end{minipage}
    \begin{minipage}[h]{0.105\textwidth}
        \centering
        \includegraphics[width=\linewidth]{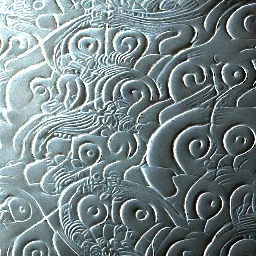}
        \includegraphics[width=\linewidth]{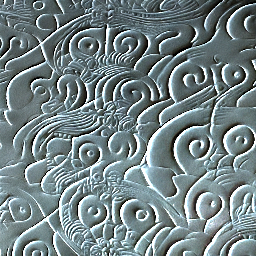}
        \includegraphics[width=\linewidth]{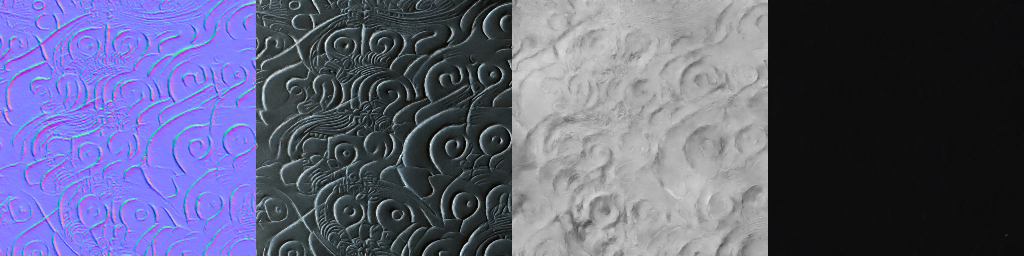}
        
        \hfill
    \end{minipage}

    \begin{minipage}[h]{0.21\textwidth}
        \centering
        \includegraphics[width=\linewidth, trim={2cm 3cm 2cm 1cm},clip]{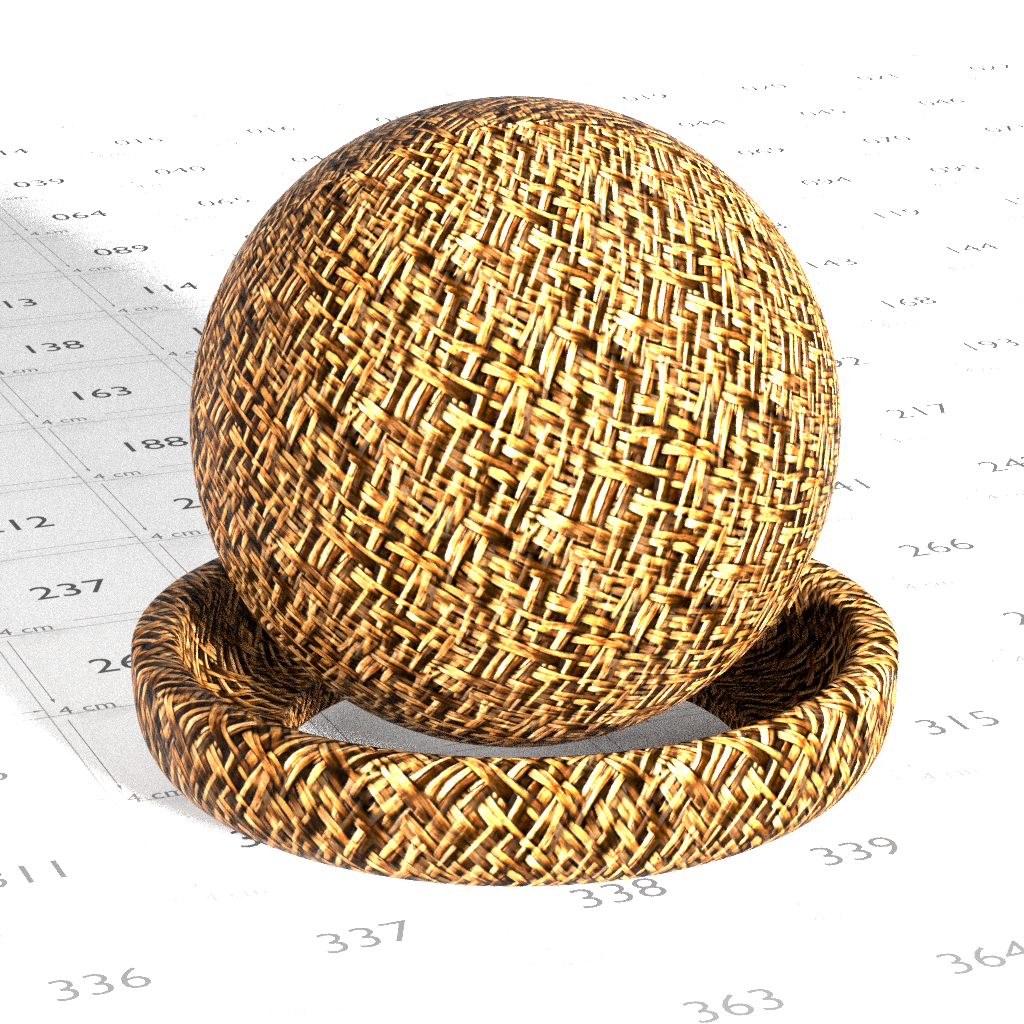}
        \input{image3/run_243/prompt.txt}
    \end{minipage}
    \begin{minipage}[h]{0.105\textwidth}
        \centering
        \includegraphics[width=\linewidth]{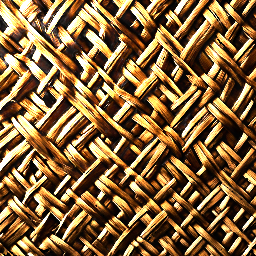}
        \includegraphics[width=\linewidth]{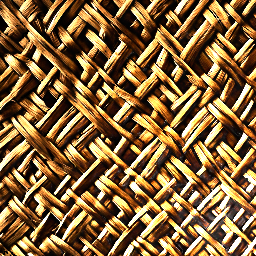}
        \includegraphics[width=\linewidth]{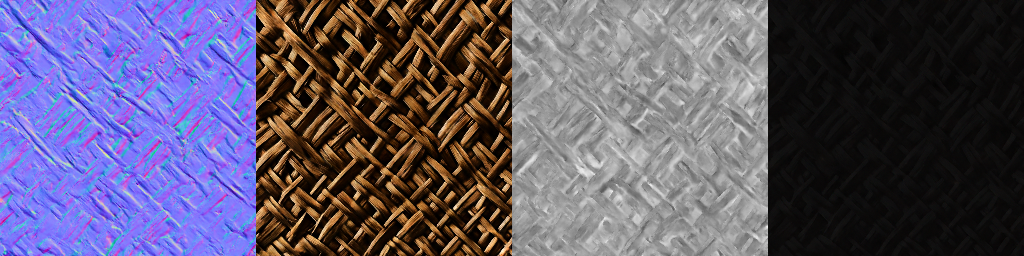}
        
        \hfill
    \end{minipage}

    \begin{minipage}[h]{0.21\textwidth}
        \centering
        \includegraphics[width=\linewidth, trim={2cm 3cm 2cm 1cm},clip]{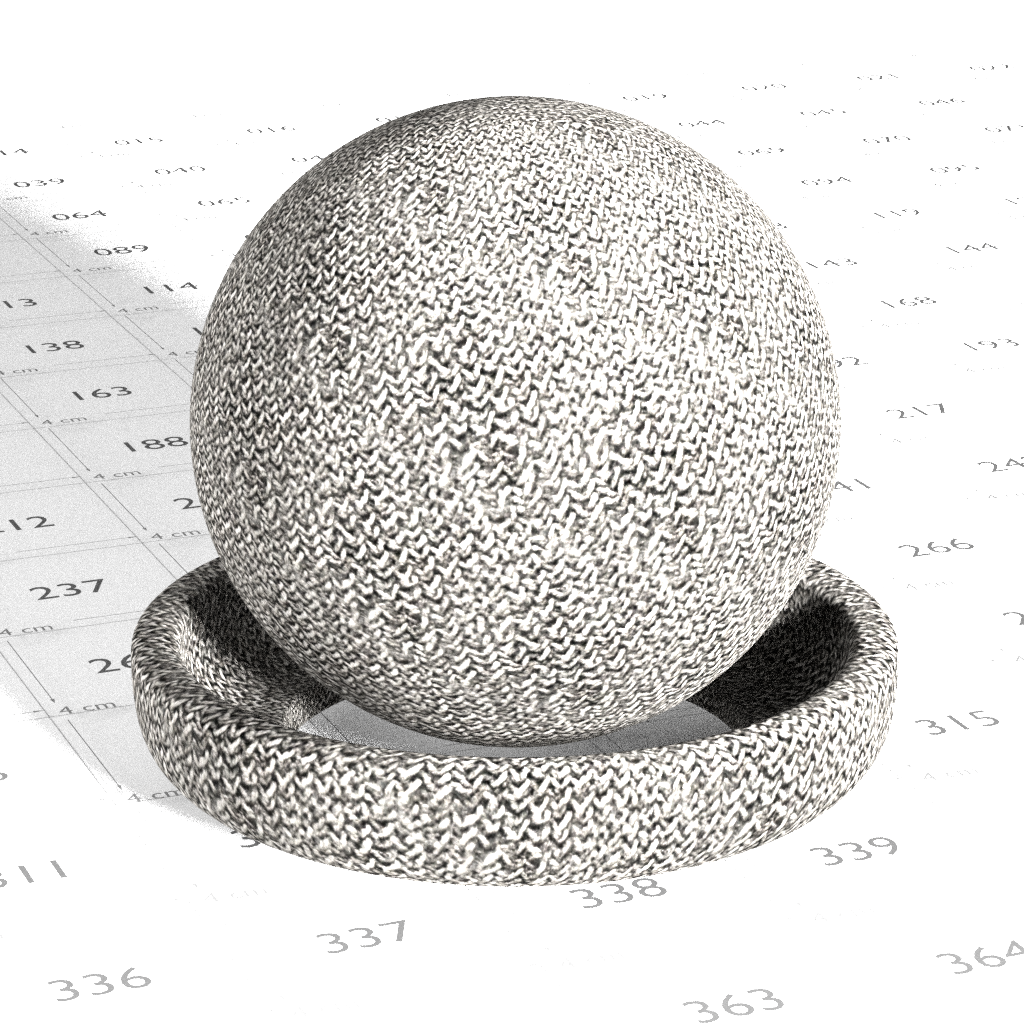}
        \input{image3/run_680/prompt.txt}
    \end{minipage}
    \begin{minipage}[h]{0.105\textwidth}
        \centering
        \includegraphics[width=\linewidth]{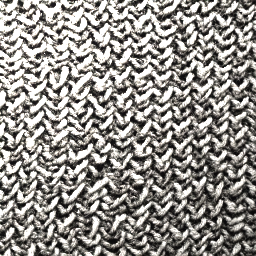}
        \includegraphics[width=\linewidth]{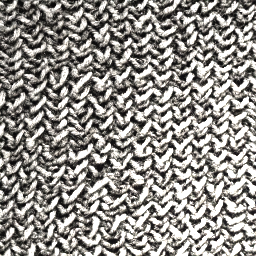}
        \includegraphics[width=\linewidth]{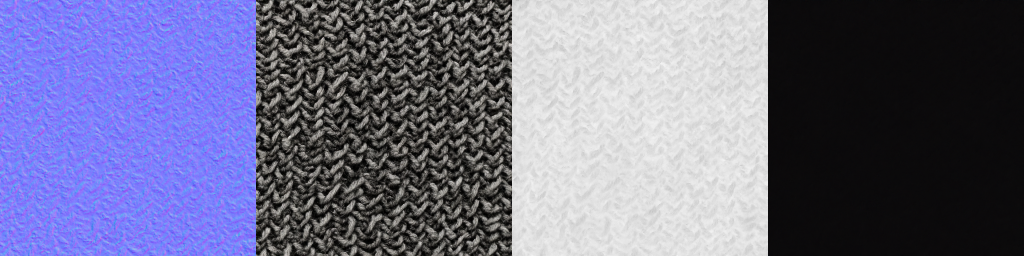}
        
        \hfill
    \end{minipage}
{tabular}
    \caption{\textbf{Stationary results.} We demonstrate our method for generating SVBRDF maps, with two flat renderings illuminated from the top left and bottom right corners shown on the right. The same texture, tiled and applied to a 3D shape, is displayed on the left.}
    \label{fig_simpleresult} 
\end{figure*}

\begin{figure*}[h!]
    \centering
    \setlength{\tabcolsep}{2pt}
    
    \begin{tabular}{ccc}

    \begin{minipage}{.33\textwidth}
        \centering
        \includegraphics[width=.325\linewidth]{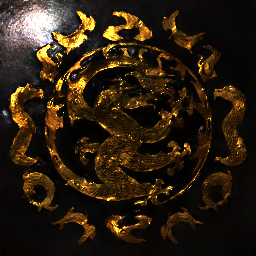}
        \includegraphics[width=.325\linewidth]{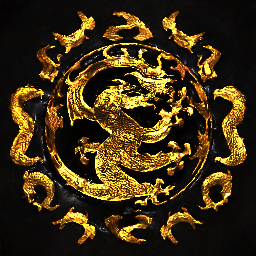}
        \includegraphics[width=.325\linewidth]{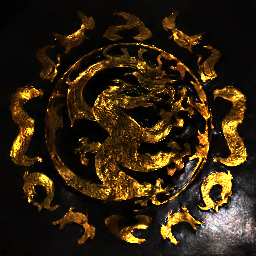}
        
        
        \includegraphics[width=0.99\linewidth]{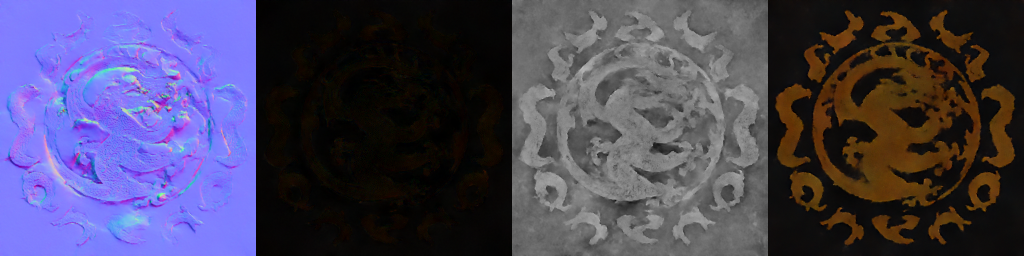}
        
        \input{image3/run_663/prompt.txt}
    \end{minipage}

    \begin{minipage}{.33\textwidth}
        \centering
        \includegraphics[width=.325\linewidth]{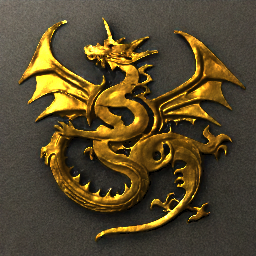}
        \includegraphics[width=.325\linewidth]{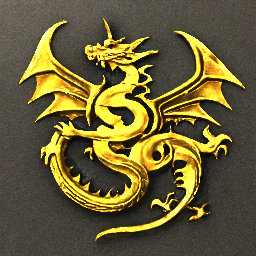}
        \includegraphics[width=.325\linewidth]{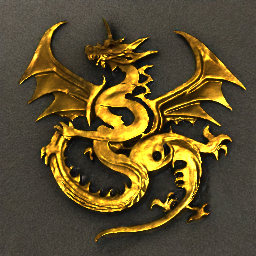}
        
        
        \includegraphics[width=0.99\linewidth]{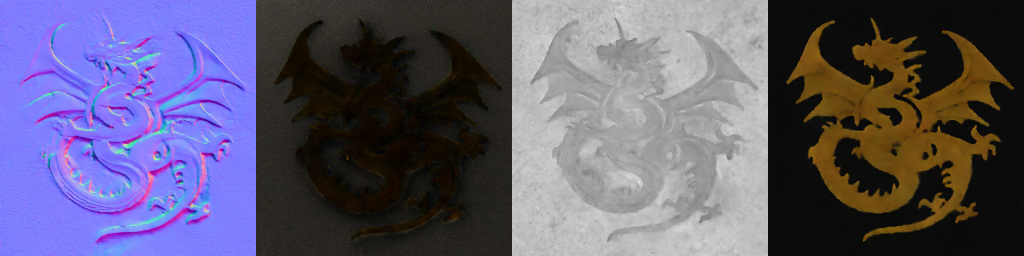}
        
        \input{image3/run_473/prompt.txt}
    \end{minipage}

    \begin{minipage}{.33\textwidth}
        \centering
        \includegraphics[width=.325\linewidth]{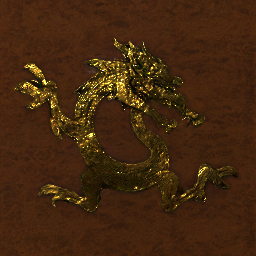}
        \includegraphics[width=.325\linewidth]{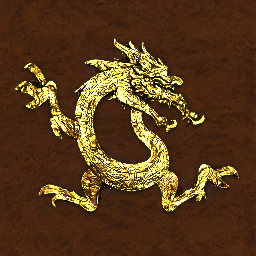}
        \includegraphics[width=.325\linewidth]{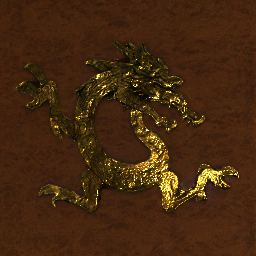}
        
        
        \includegraphics[width=0.99\linewidth]{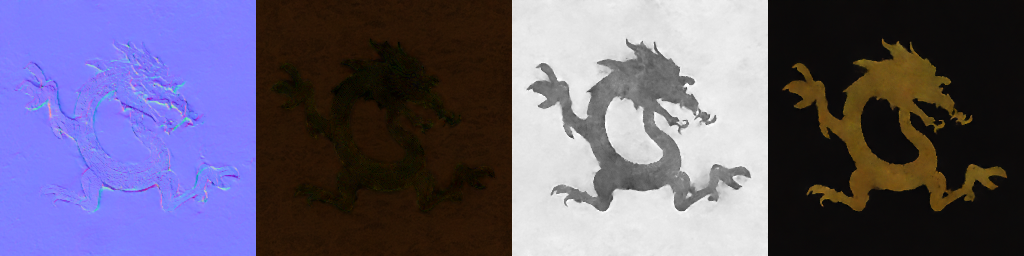}
        
        \input{image3/run_658/prompt.txt}
    \end{minipage}

    \begin{minipage}{.33\textwidth}
        \centering
        \includegraphics[width=.325\linewidth]{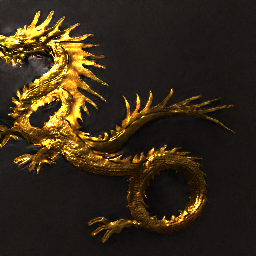}
        \includegraphics[width=.325\linewidth]{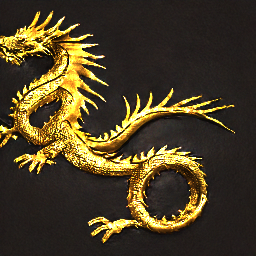}
        \includegraphics[width=.325\linewidth]{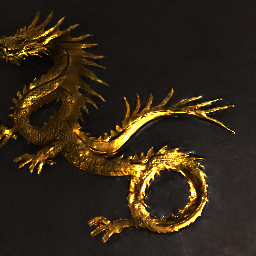}
        
        
        \includegraphics[width=0.99\linewidth]{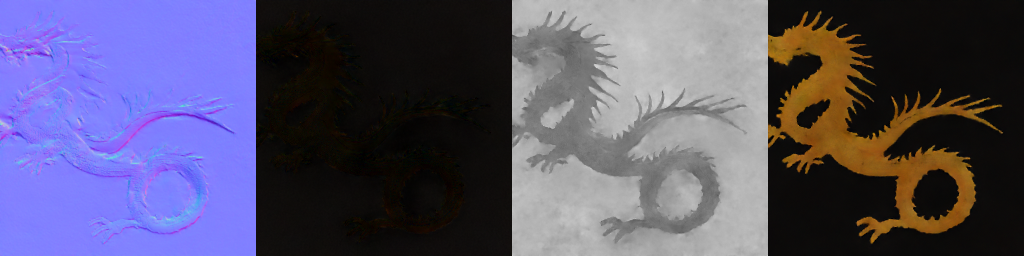}
        
        \input{image3/run_659/prompt.txt}
    \end{minipage}

    \begin{minipage}{.33\textwidth}
        \centering
        \includegraphics[width=.325\linewidth]{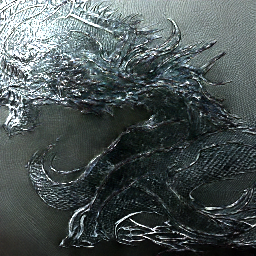}
        \includegraphics[width=.325\linewidth]{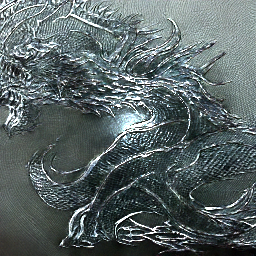}
        \includegraphics[width=.325\linewidth]{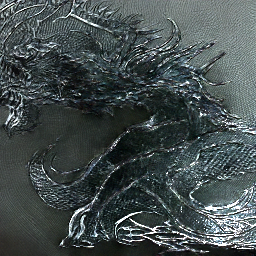}
        \includegraphics[width=0.99\linewidth]{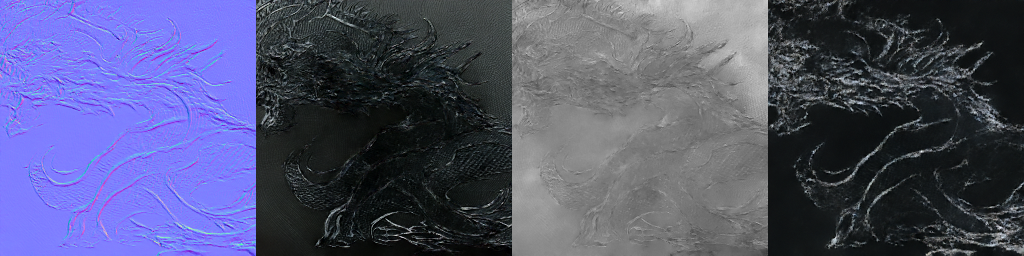}

        \input{image3/run_297/prompt.txt}
    \end{minipage}

    \begin{minipage}{.33\textwidth}
        \centering
        \includegraphics[width=.325\linewidth]{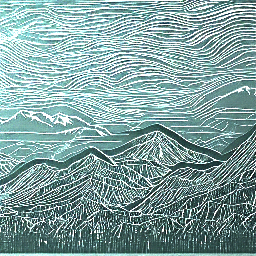}
        \includegraphics[width=.325\linewidth]{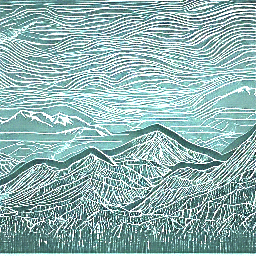}
        \includegraphics[width=.325\linewidth]{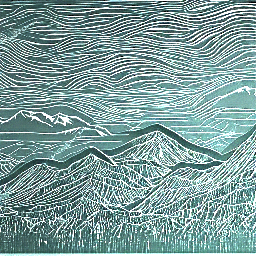}
        \includegraphics[width=0.99\linewidth]{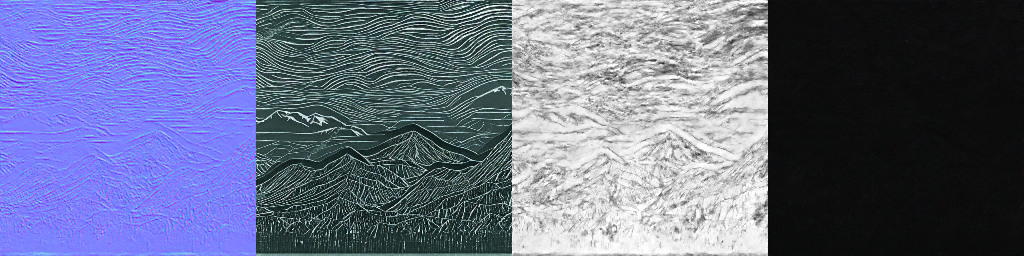}

        \input{image3/run_671/prompt.txt}
    \end{minipage}

    \begin{minipage}{.33\textwidth}
        \centering
        \includegraphics[width=.325\linewidth]{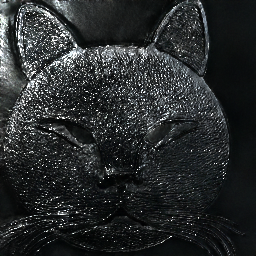}
        \includegraphics[width=.325\linewidth]{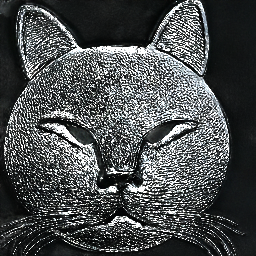}
        \includegraphics[width=.325\linewidth]{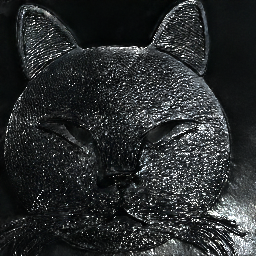}
        
        
        \includegraphics[width=0.99\linewidth]{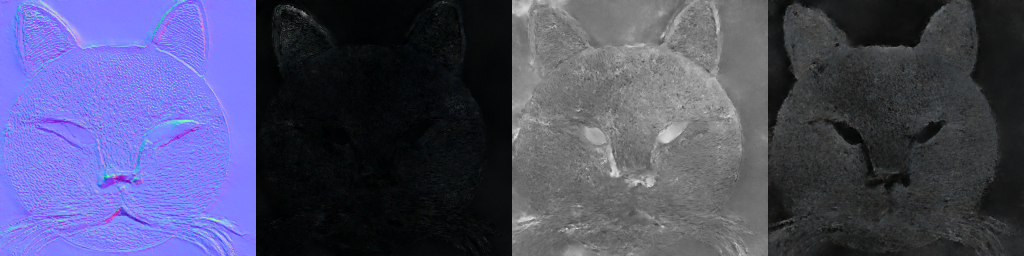}
        
        \input{image3/run_1050/prompt.txt}
    \end{minipage}

    \begin{minipage}{.33\textwidth}
        \centering
        \includegraphics[width=.325\linewidth]{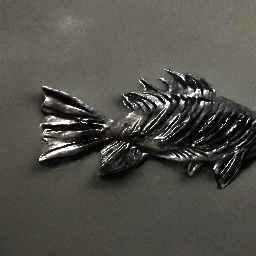}
        \includegraphics[width=.325\linewidth]{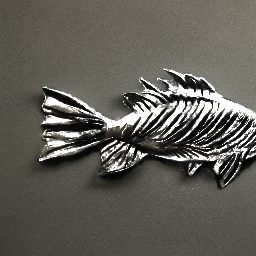}
        \includegraphics[width=.325\linewidth]{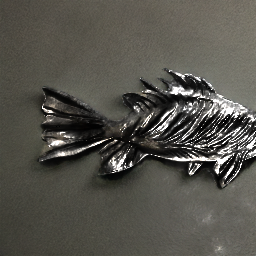}
        
        
        \includegraphics[width=0.99\linewidth]{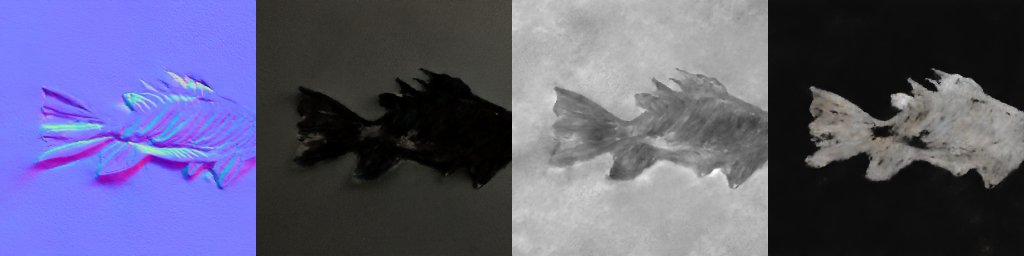}
        
        \input{image3/run_1192/prompt.txt}
    \end{minipage}

    \begin{minipage}{.33\textwidth}
        \centering
        \includegraphics[width=.325\linewidth]{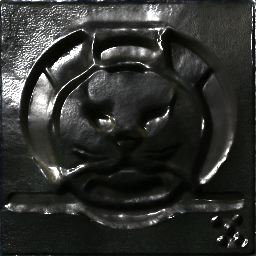}
        \includegraphics[width=.325\linewidth]{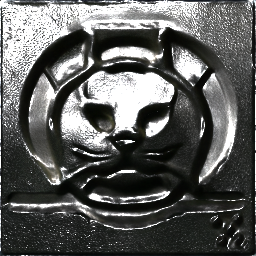}
        \includegraphics[width=.325\linewidth]{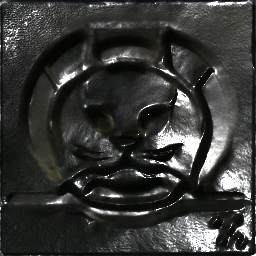}
        
        
        \includegraphics[width=0.99\linewidth]{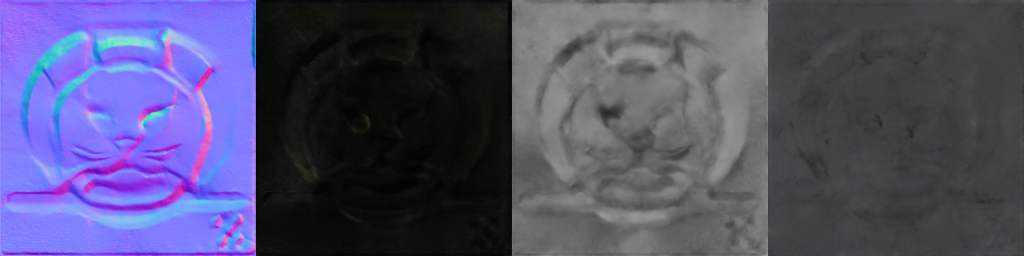}
        
        \input{image3/run_1022/prompt.txt}
    \end{minipage}



    \end{tabular}
    \caption{\textbf{Metal and glass results.} The upper half displays three flat renderings of the material, illuminated from the top left, center, and bottom right. For a continuous light interaction, please refer to the supplementary video. The lower half presents the corresponding SVBRDF maps.}
    \label{fig_metal} 
\end{figure*}

\begin{figure*}[h!]
    \centering
    \begin{minipage}[c]{.02\textwidth}
        \rotatebox{90}{\cite{guo2023text2mat}\hspace*{10em}Ours}
    \end{minipage}%
    \begin{minipage}[c]{.97\textwidth}
        \begin{subfigure}{.19\textwidth}
            \includegraphics[width=\linewidth]{image3/run_0441/00.png} 
            \includegraphics[width=\linewidth]{image3/run_0441/renderimage1d.png}
            \includegraphics[width=\linewidth]{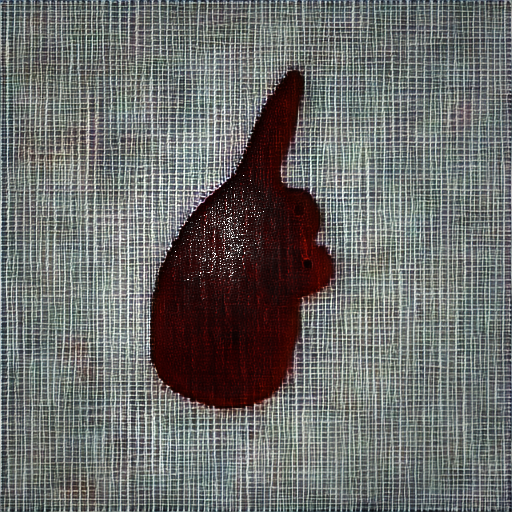}
            \includegraphics[width=\linewidth]{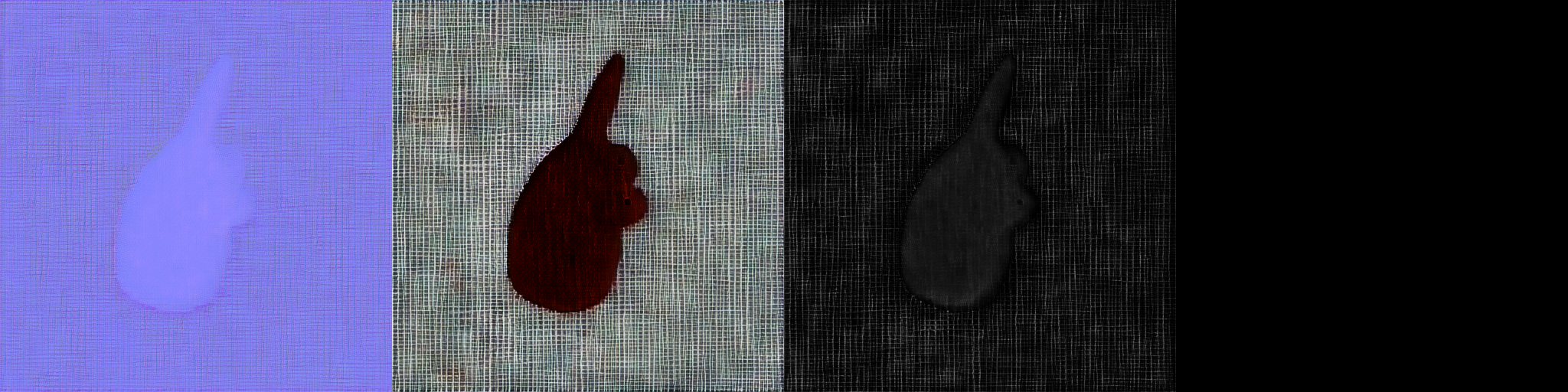}
            \caption*{rabbit carved wood}
        \end{subfigure}
        \begin{subfigure}{.19\textwidth}
            \includegraphics[width=\linewidth]{image3/run_804/00.png} 
            \includegraphics[width=\linewidth]{image3/run_804/renderimage1d.png}
            \includegraphics[width=\linewidth]{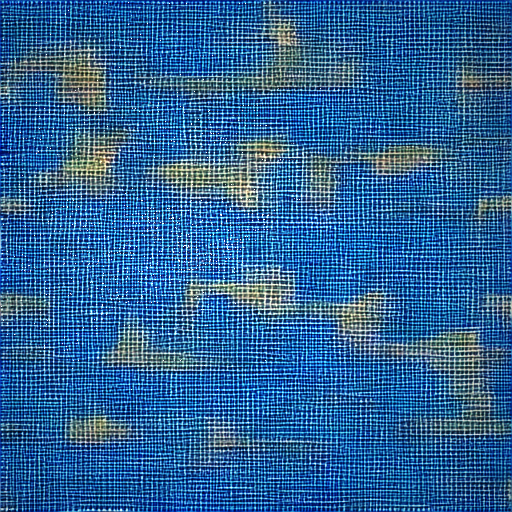}
            \includegraphics[width=\linewidth]{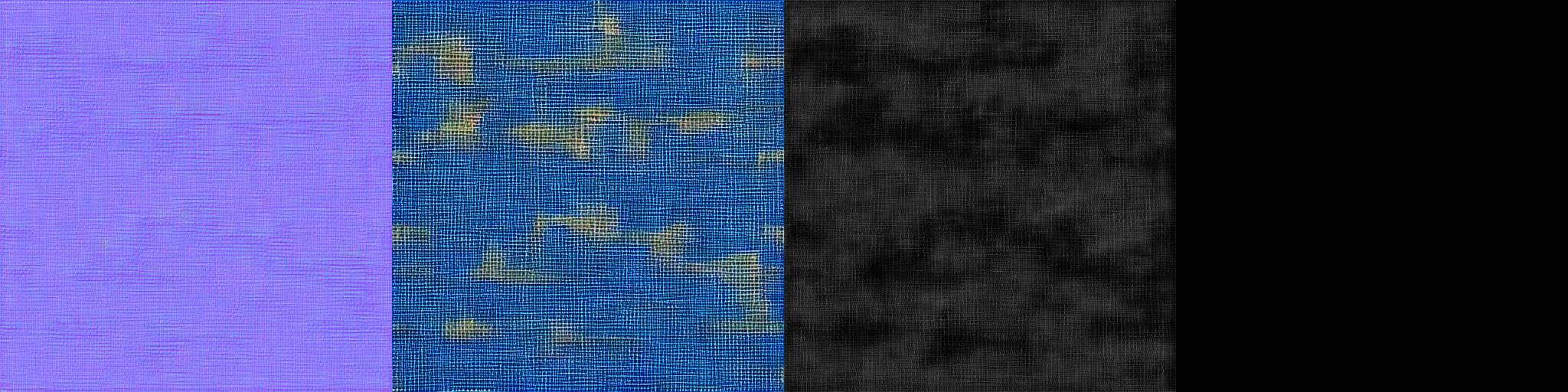}
            \caption*{bike embroidered fabric}
        \end{subfigure}
        \begin{subfigure}{.19\textwidth}
            \includegraphics[width=\linewidth]{image3/run_819/00.png}
            \includegraphics[width=\linewidth]{image3/run_819/renderimage1d.png}
            \includegraphics[width=\linewidth]{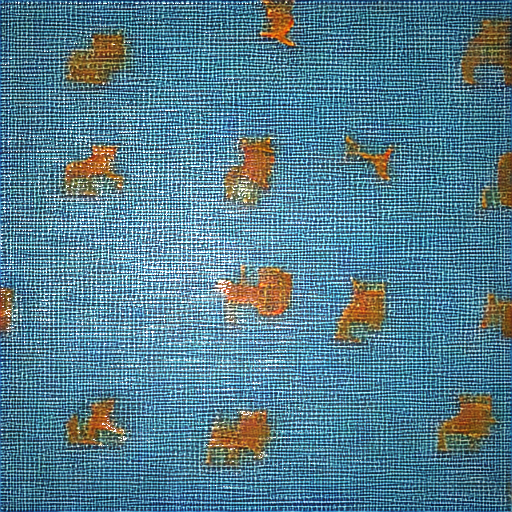}
            \includegraphics[width=\linewidth]{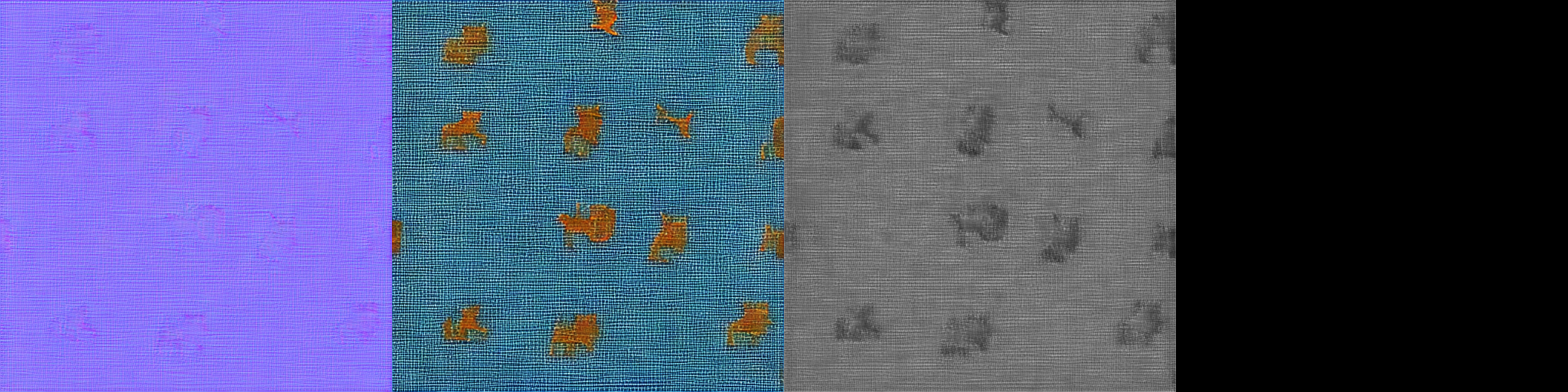}
            \caption*{cat embroidered fabric}
        \end{subfigure}
        \begin{subfigure}{.19\textwidth}
            \includegraphics[width=\linewidth]{image3/run_659/04.png}
            \includegraphics[width=\linewidth]{image3/run_659/renderimage1d.png}
            \includegraphics[width=\linewidth]{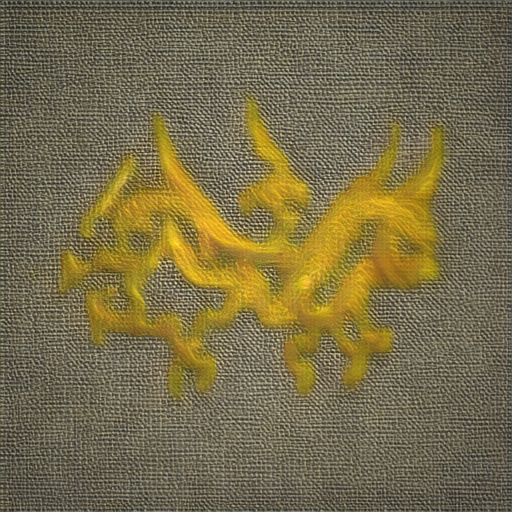}
            \includegraphics[width=\linewidth]{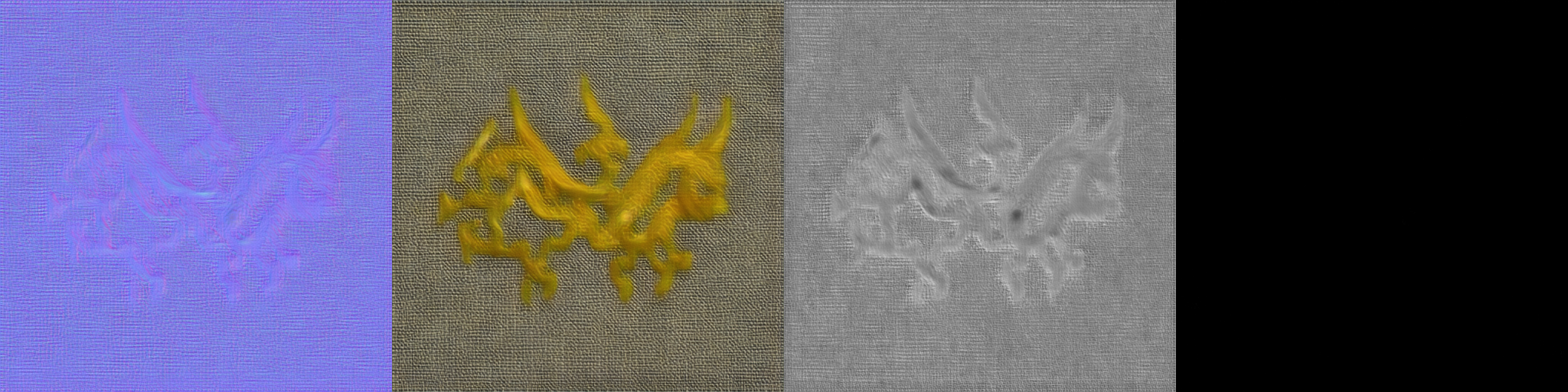}
            \caption*{dragon molded gold}
        \end{subfigure}
        \begin{subfigure}{.19\textwidth}
            \includegraphics[width=\linewidth]{image3/run_601/00.png}
            \includegraphics[width=\linewidth]{image3/run_601/renderimage1d.png} 
            \includegraphics[width=\linewidth]{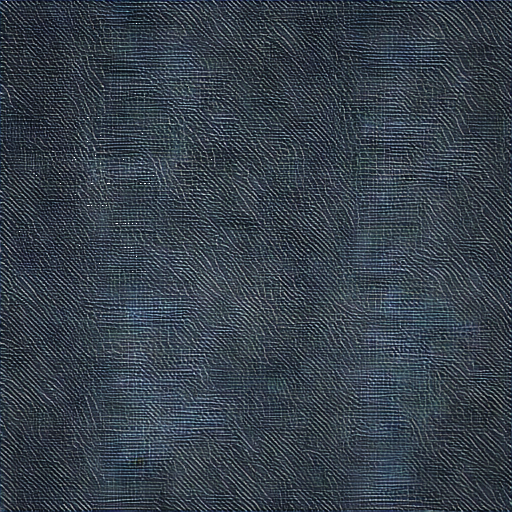}
            \includegraphics[width=\linewidth]{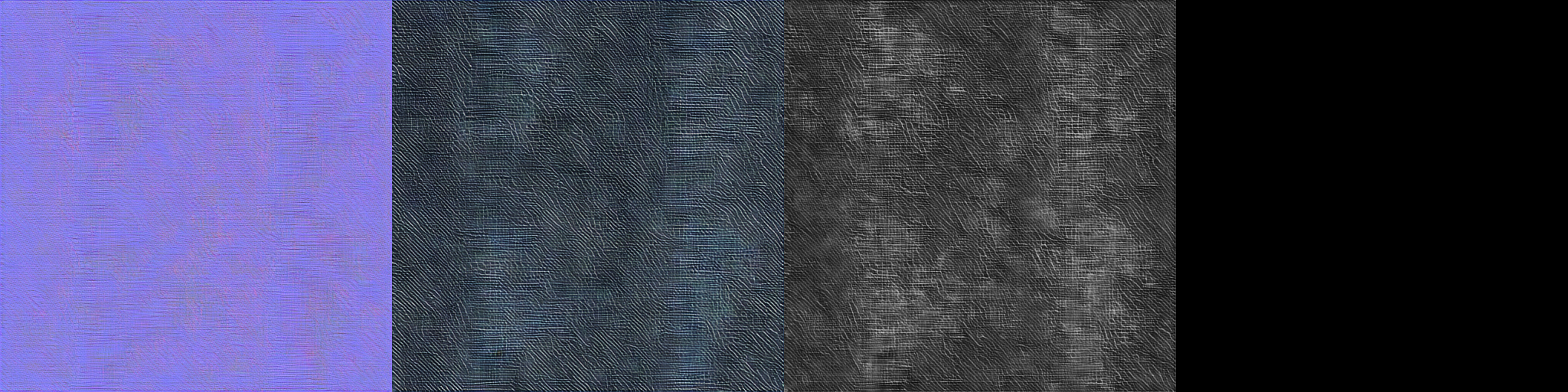}
            \caption*{dragon carved wood}
        \end{subfigure}
    \end{minipage}
    \caption{\textbf{Comparison to previous model.} The upper half displays our flat render results, with the corresponding SVBRDF maps shown beneath, compared to results from Text2Mat \cite{guo2023text2mat}. Our results exhibit a significant improvement in texture generation quality over Text2Mat, which struggles to accurately convey the meaning of the text.}
    \label{fig_compare}
    
\end{figure*}

\begin{figure*}[h!]
    \centering
    \setlength{\tabcolsep}{2pt}
    
    \begin{tabular}{cc}

    \begin{minipage}[h]{0.21\textwidth}
        \centering
        \includegraphics[width=\linewidth, trim={2cm 3cm 2cm 1cm},clip]{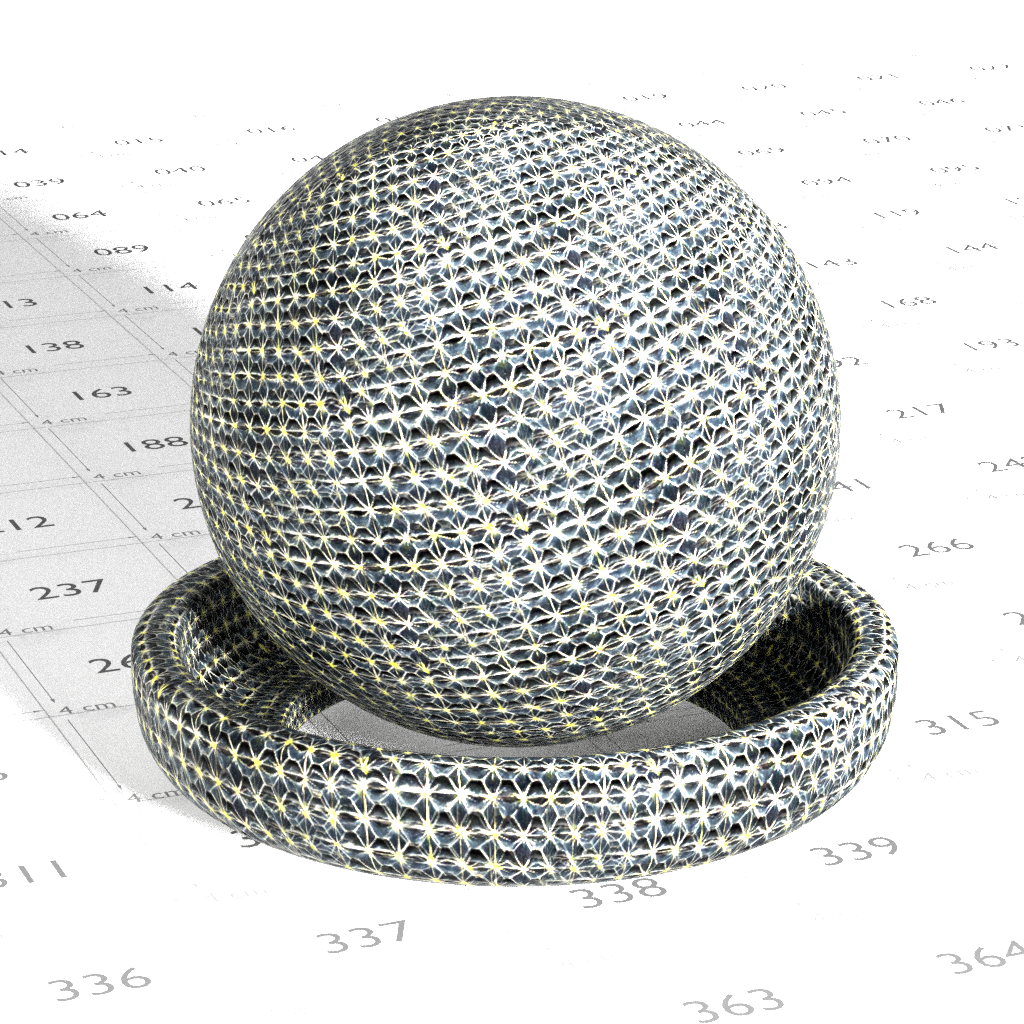}
        \input{image3/run_1215/prompt.txt}
    \end{minipage}
    \begin{minipage}[h]{0.105\textwidth}
        \centering
        \includegraphics[width=\linewidth]{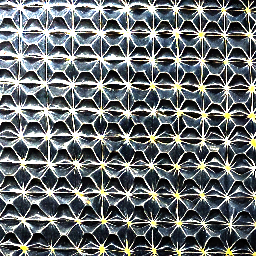}
        \includegraphics[width=\linewidth]{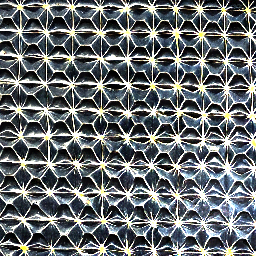}
        \includegraphics[width=\linewidth]{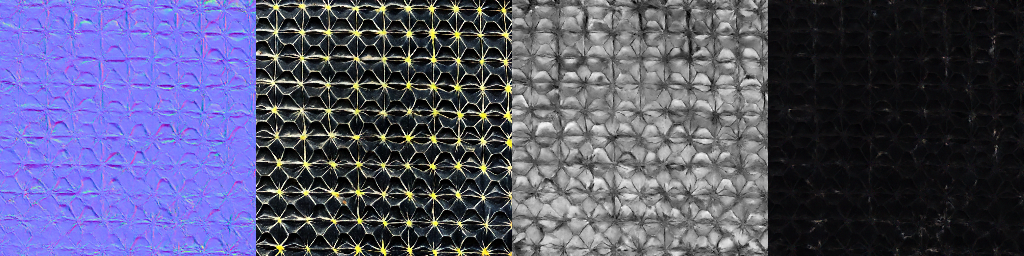}
        
        \hfill
    \end{minipage}

    \begin{minipage}[h]{0.21\textwidth}
        \centering
        \includegraphics[width=\linewidth, trim={2cm 3cm 2cm 1cm},clip]{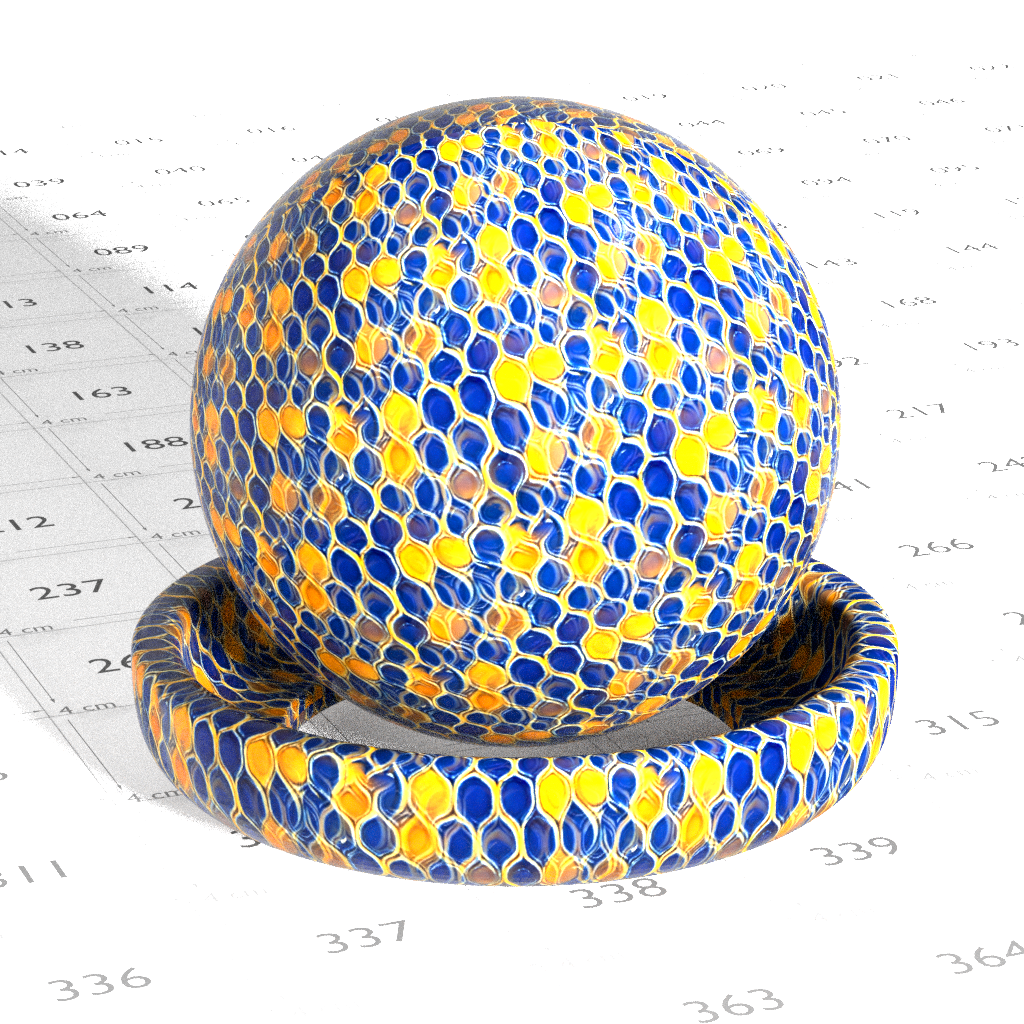}
        \input{image3/run_756/prompt.txt}
    \end{minipage}
    \begin{minipage}[h]{0.105\textwidth}
        \centering
        \includegraphics[width=\linewidth]{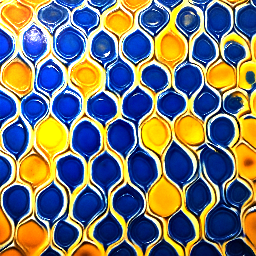}
        \includegraphics[width=\linewidth]{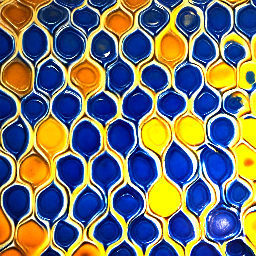}
        \includegraphics[width=\linewidth]{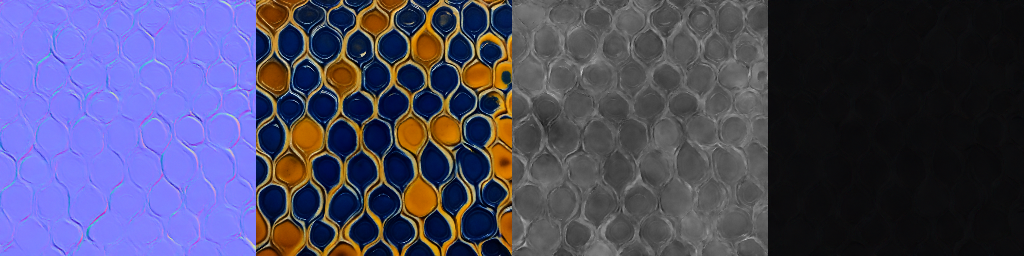}
        
        \hfill
    \end{minipage}

    \begin{minipage}[h]{0.21\textwidth}
        \centering
        \includegraphics[width=\linewidth, trim={2cm 3cm 2cm 1cm},clip]{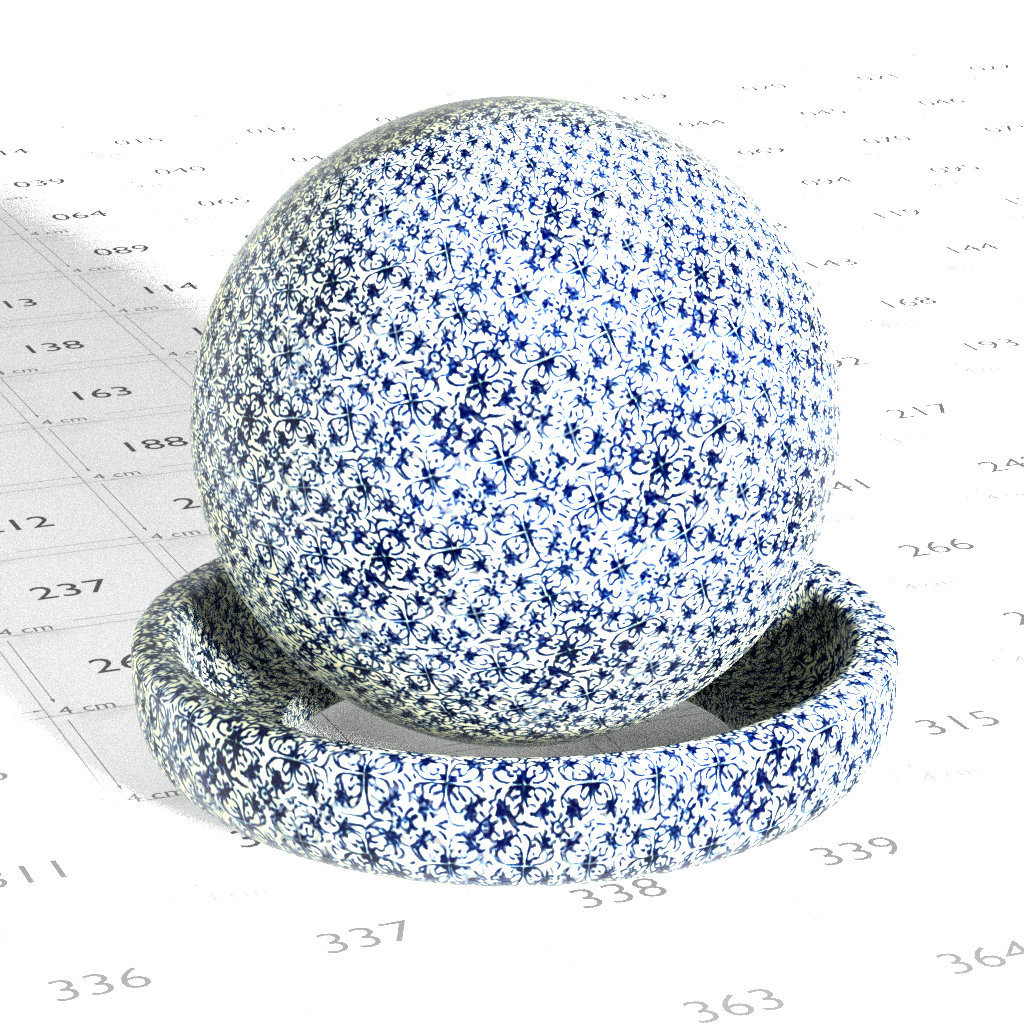}
        \input{image3/run_769/prompt.txt}
    \end{minipage}
    \begin{minipage}[h]{0.105\textwidth}
        \centering
        \includegraphics[width=\linewidth]{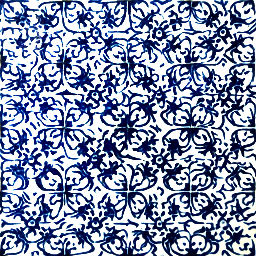}
        \includegraphics[width=\linewidth]{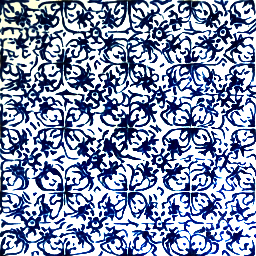}
        \includegraphics[width=\linewidth]{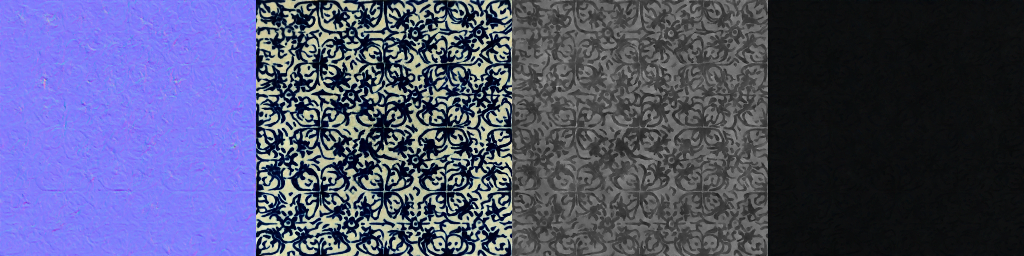}
        
        \hfill
    \end{minipage}
{tabular}
    \caption{\textbf{Editing shape and color.} We demonstrate how text can be used to control the appearance of materials, ceramic shown here, by using detailed descriptions to dictate color and pattern variations.}
    \label{fig_colorresult} 
\end{figure*}

\begin{figure*}[h!]
    \centering

    \begin{minipage}{\textwidth}
        \centering
        \includegraphics[width=0.118\textwidth]{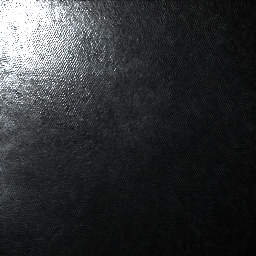}
        \includegraphics[width=0.118\textwidth]{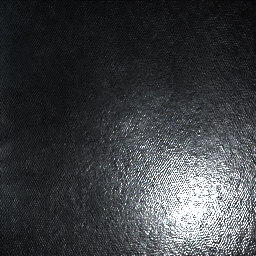}
        \includegraphics[width=0.118\textwidth]{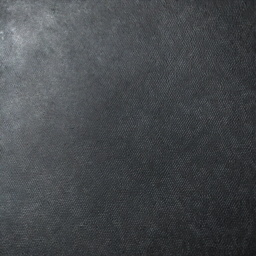} 
        \includegraphics[width=0.118\textwidth]{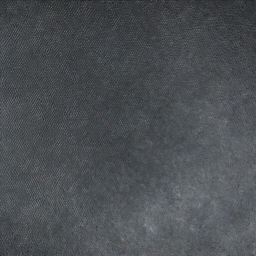}
        \includegraphics[width=0.118\textwidth]{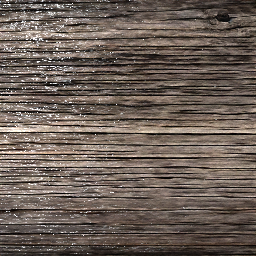} 
        \includegraphics[width=0.118\textwidth]{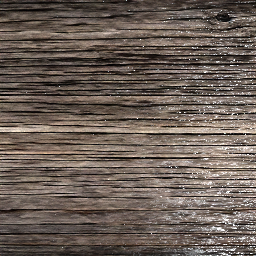} \includegraphics[width=0.118\textwidth]{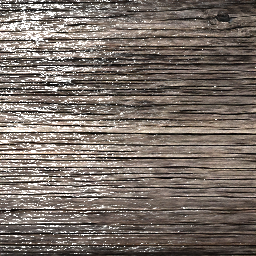}
        \includegraphics[width=0.118\textwidth]{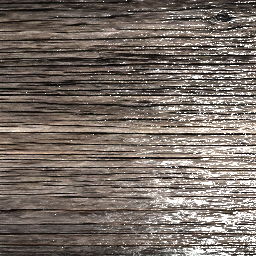}
    \end{minipage}

    \begin{minipage}{\textwidth}
        \centering
        \begin{subfigure}[b]{0.24\textwidth}
            \includegraphics[width=\textwidth,trim=255 0 0 0, clip]{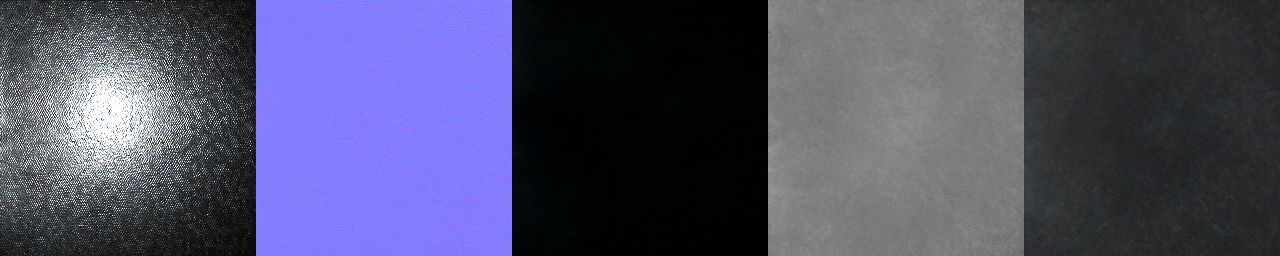}
            \caption{metal with low roughness}
        \end{subfigure}
        \begin{subfigure}[b]{0.24\textwidth}
            \includegraphics[width=\textwidth,trim=260 0 0 0, clip]{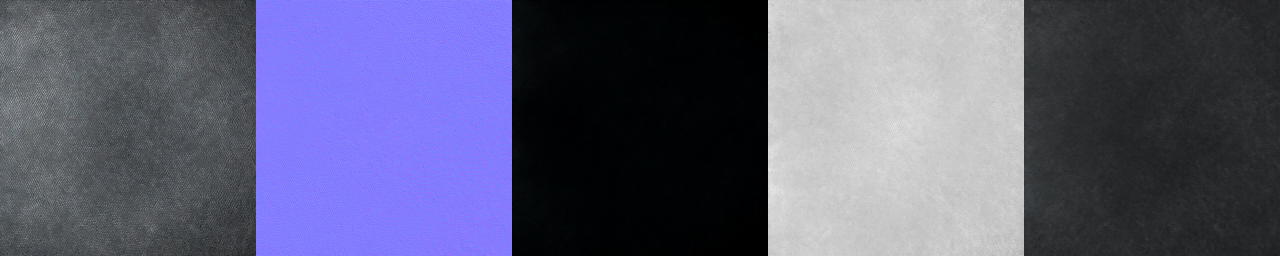}
            \caption{metal with high roughness}
        \end{subfigure}
        \begin{subfigure}[b]{0.24\textwidth}
            \includegraphics[width=\textwidth,trim=260 0 0 0, clip]{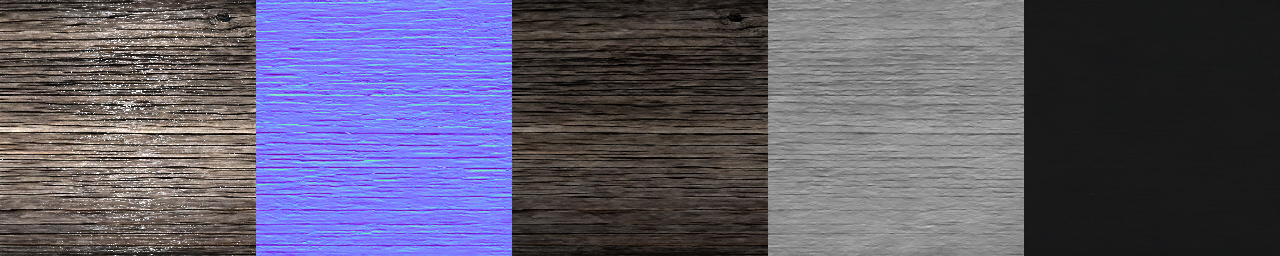}
            \caption{wood with low specular}
        \end{subfigure}
        \begin{subfigure}[b]{0.24\textwidth}
            \includegraphics[width=\textwidth,trim=260 0 0 0, clip]{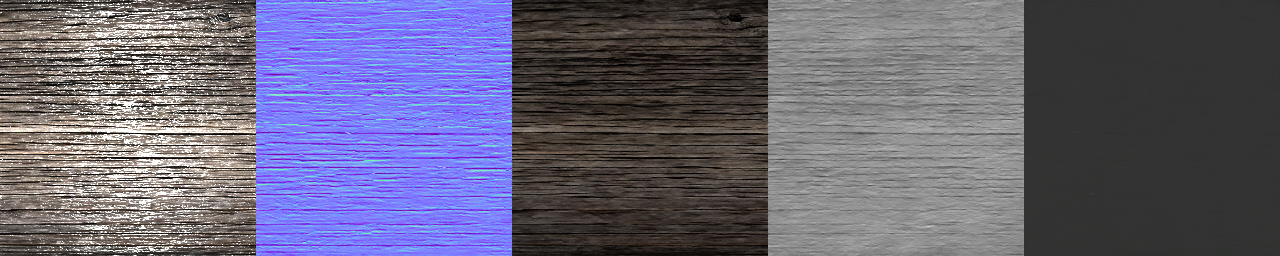}
            \caption{wood with high specular}
        \end{subfigure}
    \end{minipage}

    \caption{\textbf{Controlling physical parameters with text.} We show two flat renderings illuminated from the top left and top right corners, with corresponding SVBRDF maps displayed below. The metal example demonstrates roughness adjustment for the same sample by modifying the text prompt, and the wood example illustrates control over specularity.}
    \label{fig_specular}
\end{figure*}


\section{Implementation}
\subsection{Dataset and Training}
To train our network, we utilized the SVBRDF dataset introduced by INRIA synthetic SVBRDF dataset \cite{deschaintre2018single} and UBO dataset \cite{merzbach2020bonn}, which comprises approximately 200,000 SVBRDFs alongside their corresponding rendered images. These datasets encapsulate a wide array of material types such as leather, fabric, and stone, with the specular maps employing Schlick's approximation for reflection modeling. The grey blocks shown in \ref{fig_architecture} are used as pre-trained while the rest are fully trained using our dataset. During the training of this network, it initially extracts rendered images from the dataset. These images are then processed separately through a VAE network and the initial layers of a VGG network to extract features. Subsequently, the obtained latent representations are reshaped and used as inputs for the network. Throughout the training, noise is incrementally introduced into the SVBRDF maps until it morphs into Gaussian noise. Moreover, the training dynamically assesses the necessity of inputting roughness and specular parameters into the network; if required, these parameters are calculated based on the specular and roughness maps and then fed as inputs. We employed a standardized input format of "flat texture of + text". We omitted the repeated phrase "flat texture of" in all results for brevity. Our training process was conducted on four A100 GPUs, utilizing FP16 precision, with each training cycle lasting approximately ten days.


\subsection{Performance}
In conducting 100 network queries on an A100 GPU and averaging the time expenditure, Stable Diffusion takes 3.21 seconds to produce a static rendered image, whereas our architecture requires 5.16 seconds to be queried and generate the four SVBRDF maps. Although our approach is marginally slower than models dedicated to direct image generation, the disparity is minimal. The advantage gained, however, is the provision of fully editable and freely utilizable SVBRDF maps instead of single images. 

\subsection{Experiments}
Initially, we experimented with three approaches, the results of which are presented in an ablation study illustrated in Fig. \ref{fig:ablation}. The first strategy was to transfer learning from the Stable Diffusion 2 \cite{rombach2022high} model, starting by training its VAE. Given our goal to ensure model versatility, our first idea was to freeze the encoder part of the VAE's model, train the VAE's decoder, adjust it to output ten channels, increase the network layers, and train with the SVBRDF dataset. After the VAE was trained and its performance deemed acceptable, we proceeded to train the entire network end-to-end with the SVBRDF dataset. However, after sufficient training convergence, the network's performance was poor, producing unrealistic SVBRDF. Two examples are shown in Fig. \ref{fig:ablation}.a demonstrate that the model was unable to distinguish wood material from metal due to color specular maps. The second method we attempted was to directly train a single diffusion model for text-to-SVBRDF conversion was hindered by lacking the training dataset, resulting in a model that could only represent a limited range of specific textures and demonstrated poor text description comprehension. For instance, as depicted in Fig. \ref{fig:ablation}.(b), it failed to generate materials not present in the dataset, such as a carved bunny, since the model was only trained on plain wood. The final approach, which we currently employ, involves using a two-phase diffusion model pipeline shown in \ref{fig:ablation}.(c).

\section{Results}

\subsection{Main Results}

\par\textbf{Non-stationary:}
 In Fig. \ref{fig_mainresult}, we demonstrate our method for generating SVBRDF maps followed by two flat renderings illuminated from the top left and bottom right corner. On the left, we show the same texture applied to a 3D shape. It is observable that our model is capable of producing meaningful textures following the text, such as bike-embroidered fabric and rabbit-carved wood; which presents certain challenges for artists. Furthermore, we can guide the texture generation with more complex language instructions, like glazed ceramic with blue and white, allowing us to control both the material and color. In order to seamlessly tile the textures, we synthesize a gradient mask to blend the original and quarter-shifted versions of the image. This tiling algorithm facilitates the creation of a continuous pattern that is suitable for repetitive tiling applications.

\par\textbf{Stationary:}
In Fig. \ref{fig_simpleresult}, we have generated stationary results including leather, brick, wood, rubber, fabric, and stone. The rubber outcome has a diffuse map that is almost entirely black and is primarily dominated by the specular map, showcasing our network's capability to generate challenging materials accurately without baking the highlights in the maps.

\par\textbf{Metal and glass:}
In Fig. \ref{fig_metal}, we use text to control the generation of metal and glass materials. The metal materials have colored specular maps unlike other results, and the rendered results exhibit metallic characteristics. In this figure, the three rendered images show the light source moving from the top left to the bottom right. Additionally, we have also generated glass materials that present a dielectric feeling. For simplicity, we chose to only show the flat renderings of metal and glass examples with the absence of light transmission.

For additional results and detailed analyses, please refer to the supplementary materials.

\subsection{Comparison to Previous Work}
To our knowledge, the only existing model for converting text to materials is "Text2Mat: Generating Materials from Text." \cite{guo2023text2mat}. We compared our results with theirs, as shown in Fig. \ref{fig_compare}, where the text input into both models is provided at the bottom. It can be seen that Text2Mat has difficulty creating meaningful textures; it can only generate basic stationary materials and cannot fulfill complex requirements. Since Text2Mat is not open source, these results were obtained by sharing our text directly with the authors and requesting generation, which we would like to appreciate.

\subsection{Editing Results}
\par\textbf{Text-controlled appearance features:}
Our model can manipulate the appearance features of the generated SVBRDF map, such as color and pattern, using text prompts. For single text input, our two-stage diffusion-based methods can produce multiple different textures for user to select. As illustrated in Fig. \ref{fig_colorresult}, we can control the texture's color or local shape through text by keeping the material constant while varying the detailed descriptions.

\par\textbf{Text-Controlled physical parameters:}
Our network can also control physical parameters through text, specifically roughness, and specular levels. In Fig. \ref{fig_specular}, we demonstrate our method for controlling roughness and specular followed by two flat renderings illuminated from the top left and bottom right corner. By using a text controller, we can adjust these parameters simply by inputting text commands like "high/low roughness/specular" to manipulate the output's roughness and specular intensity, marked as Text 2 in Fig. \ref{fig_architecture}. This can be also done by post-processing the generated SVBRDF maps that allow further editing of the physical parameters. 

\section{Discussion and Conclusion}
\subsection{Limitation and future work}
Our model occasionally bakes highlights into the diffuse map. Furthermore, because diffusion occurs directly in image space, the resolution of our model is limited and currently low. This issue might be addressed by exploring latent-based diffusion. Similarly to other diffusion models, our model sometimes generates irrelevant or nonsensical results relative to the input text. However, we believe these issues can be overcome with further research.
\subsection{Conclusion}
Our model achieves generation from text to material, capable of producing specified textures or colors, thus filling a gap in related research. Furthermore, we can also create materials with special semantics, such as a Rabbit carved wood. To the best of our knowledge, we are the only ones able to perform this type of generation. Lastly, we would like to thank the organizations that provided the UBO and INRIA synthetic SVBRDF datasets and express our gratitude to Text2Mat for providing comparison images.

\printbibliography
\end{document}